\documentclass[aps,prb,twocolumn,superscriptaddress]{revtex4-2}
\usepackage{amsmath,braket,txfonts}
\usepackage[pdftex]{graphicx}
\usepackage[pdftex,colorlinks=true,bookmarks=true,citecolor=blue,linkcolor=blue,urlcolor=blue,breaklinks=true]{hyperref}

\begin{document}
\title{Dynamical conductivity of disordered quantum chains}
\author{Shintaro Takayoshi}
\affiliation{Department of Physics, Konan University, Kobe 658-8501, Japan}
\author{Thierry Giamarchi}
\affiliation{Department of Quantum Matter Physics, University of Geneva,
Geneva 1211, Switzerland}

\date{\today}

\begin{abstract}
We study the transport properties of a one dimensional quantum system
with disorder. We numerically compute the frequency dependence of the conductivity
of a fermionic chain with nearest neighbor interaction
and a random chemical potential by using the Chebyshev matrix product state 
(CheMPS) method.
As a benchmark, we investigate the noninteracting case first.
Comparison with exact diagonalization and
analytical solutions demonstrates that the results of CheMPS are reliable over a wide range of frequencies. 
We then calculate the dynamical conductivity spectra of the interacting system for various values of the interaction and disorder strengths. 
In the high frequency regime, the conductivity decays as a power law, with an interaction dependent exponent.  
This behavior is qualitatively consistent with the bosonized field theory predictions, although the numerical evaluation of the 
exponent shows deviations from the analytically expected values. We also compute the characteristic pinning frequency at which a peak 
in the conductivity appears. We confirm that it is directly related to the inverse of the localization length, even in the interacting case.
We demonstrate that the localization length follows 
a power law of the disorder strength with an exponent dependent on the interaction, and find good quantitative agreement with the field theory predictions. 
In the low frequency regime, we find a behavior consistent with 
the one of the noninteracting system  $\omega^{2}(\ln\omega)^{2}$ independently of the interaction.
We discuss the consequences of our finding for experiments in cold atomic gases. 
\end{abstract}

\maketitle

\section{Introduction}

Disorder has profound effects on quantum systems, 
as directly evidenced in the celebrated 
Anderson localization~\cite{Anderson1958PR}. 
As shown by Anderson, disorder can change the plane waves of free particles 
to exponentially localized states with spectacular consequences for transport. 
Anderson localization is relevant for a host of systems 
ranging from condensed matter to classical waves. 
In particular, cold atomic gases, due to their remarkable controllability, 
have been instrumental in evidencing the localization of the wavefunction 
through seminal experiments of the groups of 
A. Aspect with laser speckles~\cite{Aspect_loc2007PRL,Aspect_loc2008Nature} 
and M. Inguscio with 
quasi-periodic potentials~\cite{Inguscio_loc2007PRL,Inguscio_loc2008Nature}. 
After sixty years since its proposal, 
Anderson localization still continues to present new challenges 
and mathematical developments~\cite{Filoche_landscape2012PNAS}. 

The combination of disorder and interactions poses 
an additional layer of challenge especially 
in the context of condensed matter physics. 
This complicated problem in thermalized systems was tackled 
by perturbative~\cite{AltshulerAronovLee1980PRL}, 
or renormalization group (RG) techniques 
in one dimension~\cite{GiamarchiSchulz1988PRB} 
and two dimensions~\cite{Castellani1984PRB,Finkelstein1984ZPB,BelitzKirkpatrick1994RMP}. 
While the disorder basically decelerates particles, 
which leads to a reinforcement of interactions, 
it can also weaken them due to the exponentially small overlap 
between two localized states. 
This competition is highly nontrivial and constitutes 
an intensively studied topic, 
which we call the problems of localization of interacting particles. 
Another direction related with this problem is 
to study the thermalization and ergodicity in isolated quantum systems 
with disorder and interactions, which is known under the name of many-body 
localization~\cite{GornyiMirlinPolyakov2005PRL,BaskoAleinerAltshuler2007PRB,OganesyanHuse2007PRB,Abanin_mbl2019RMP}. 

Our target in this paper is one-dimensional systems, 
where strong quantum fluctuations lead to special states 
such as the Tomonaga-Luttinger liquid (TLL) characterized by 
correlations decaying in power law~\cite{Giamarchi2004Book}, 
and the effects of interactions are particularly strong.
Furthermore, the disorder effects are also at their maximum, 
and even an infinitesimal disorder localizes 
all states for noninteracting particles. 
Thus one expects a severe competition between disorder and interactions. 
Renormalization group (RG) analysis shows the existence of 
a localized-delocalized transition 
both for fermions and bosons~\cite{GiamarchiSchulz1988PRB}. 
The localized phase for bosons, the Bose glass, 
persists in higher dimensions~\cite{Fisher_bosonloc1989PRB}, 
and cold atomic systems again provides a controlled experimental access, 
confirming the Bose glass phase in biperiodic 
systems~\cite{Modugno_bg2013PRL,Modugno_bg2014PRL}. 

To characterize disordered systems, transport is an important property. 
RG can access DC conductivity down to the temperature 
related to the inverse localization scale~\cite{GiamarchiSchulz1988PRB}. 
Below this scale, more phenomenological calculations predict 
the Mott variable range hopping behavior 
in the thermalized case of localization of interacting 
particles~\cite{Mott1970PhilMag,Nattermann2003PRL} 
and zero DC conductivity in the isolated case of many-body 
localization~\cite{AleinerAltshulerShlyapnikov2010NatPhys}. 
AC transport also reflects the competition 
between disorder and interactions. 
Dynamical conductivity is exactly known for 
Anderson localization~\cite{Berezinskii1974JTEP,Gogolin1978PSS,Gogolin1982PhysRep}. 
At high frequency, 
the behavior for the interacting particles can be extracted 
from the RG \cite{GiamarchiSchulz1988PRB}, 
and the low frequency behavior has been investigated 
by approximate methods such as 
a variational approach~\cite{GiamarchiLeDoussal1996PRB}.
Despite these efforts, no general methods are applicable 
to the full frequency range for dynamical conductivity. 
This situation is regrettable 
since cold atoms would be perfect systems to investigate 
such AC behavior of the conductivity 
with methods such as phase shaking of the optical 
lattice~\cite{TokunoGiamarchi2011PRL,Wu_shakeconduct2015EPL,Anderson_optlattconduct2019PRL}. 
Indeed in biperiodic lattices, signatures of the localization 
such as a localization peak in the amplitude shaking of the optical lattice 
have been predicted~\cite{Orso_optlattdisorder2009PRA} 
and observed~\cite{Modugno_bg2014PRL}. 

In the present paper, we study the dynamical conductivity, 
i.e., the AC transport property as a function of frequency, 
in simple spinless fermion chains 
with nearest neighbor interactions in a random chemical potential. 
We perform numerical calculations using a variant of 
a Density Matrix Renormalization Group (DMRG) method 
to compute the dynamical quantity of disordered quantum systems 
with good precision. 
We compare the obtained result for dynamical conductivity 
with the field theory, and discuss its AC behavior 
over the full frequency regime. 
Such fermionic systems can be mapped either 
to Ising anisotropic spin chains 
in a random magnetic field or to hard core bosonic chains 
in a random chemical potential. 
Thus the system on which we focus is quite generic 
to demonstrate the applicability of our method 
and to study the physics of the dynamical conductivity 
in one-dimensional disordered quantum systems. 

This paper is organized as follows.
In Sec.~\ref{sec:model}, we introduce the model of spinless fermions with a random chemical potential and nearest neighbor interactions, which is a target of this paper. This model is connected to the XXZ spin chain with a random magnetic field. We also give the expression of dynamical conductivity using the Kubo formula. 
Section~\ref{sec:methods} explains the numerical technique called Chebyshev matrix product state (CheMPS), which we mainly utilize to investigate the interacting system.
In Sec.~\ref{sec:results}, we describe the calculated results of the dynamical conductivity obtained by the numerics and its behavior in the high and low frequency regimes. The numerical results are compared with analytical prediction from field theory.
We summarize our results and discuss future problems
in Sec.~\ref{sec:summary}.

\section{Model and physical quantities} \label{sec:model}

Let us consider a spinless fermion system 
with a nearest-neighbor interaction 
\begin{align}
 \hat{\mathcal{H}}=&J\sum_{l=1}^{N-1}
   \Big[\frac{1}{2}
   (\hat{a}_{l}^{\dagger}\hat{a}_{l+1}+\mathrm{H.c.})
   +\Delta \Big(\hat{n}_{l}-\frac{1}{2}\Big)
     \Big(\hat{n}_{l+1}-\frac{1}{2}\Big)\Big]
\nonumber\\
   &-\sum_{l=1}^{N}h_{l}\Big(\hat{n}_{l}-\frac{1}{2}\Big),
\label{eq:HamilrandomFermi}
\end{align}
where $N$ is the number of sites, 
$\hat{a}_{l}^{\dagger}(\hat{a}_{l})$ is 
the fermion creation (annihilation) operator at the site $l$, 
and $\hat{n}_{l}\equiv \hat{a}_{l}^{\dagger}\hat{a}_{l}$ is the number operator. 
The random chemical potential $h_{l}$ 
on each site distributes uniformly 
in the finite interval $h_{l}\in [-W,W]$. 
Thus $W$ represents the strength of disorder. 
In view of the numerical solution of this model we assume open boundary conditions, while the analytic solutions are usually performed with periodic boundary conditions.  
Without disorder,
this system is particle-hole symmetric 
and half-filled. 
In this case, the system is known to be described at low energy by a TLL Hamiltonian (see Appendix~\ref{ap:bosonization}). 
Since this model is solvable by Bethe ansatz, the TLL parameters can be exactly computed. For example, the parameter $K$ controlling the decay of 
correlation functions~\cite{Giamarchi2004Book} is given by
\begin{align}
 K=[2(1-\arccos(\Delta)/\pi)]^{-1}.
\label{eq:TLLparameter}
\end{align}

%%%%%%%%%% Fig : K(Delta) %%%%%%%%%%
\begin{figure}[t]
\centering
\includegraphics[width=0.25\textwidth]{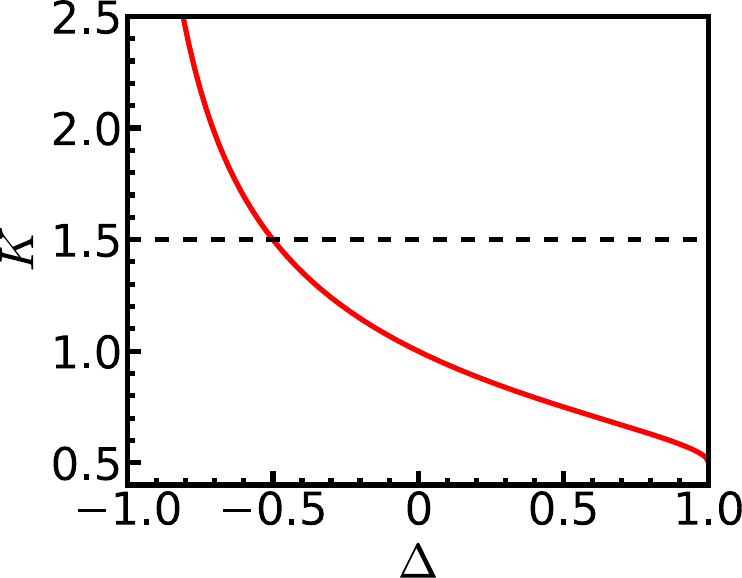}
\caption{The Solid line represents the Luttinger parameter $K$ as a function of $\Delta$ for a bosonized effective Hamiltonian of Eq.~\eqref{eq:HamilrandomFermi} without disorder. The horizontal dashed line is $K=3/2$, below which the system is localized by introducing the disorder.}
\label{fig:DelvsK}
\end{figure}
%%%%%%%%%%%%%%%%%%%%

The Jordan-Wigner transformation: 
\begin{align}
\begin{split}
 &\hat{S}_{l}^{z}=\hat{n}_{l}-\frac{1}{2},\\
 &\hat{S}_{l}^{+}=\hat{a}_{l}^{\dagger}\prod_{j=1}^{l-1}(1-2\hat{n}_{j}),\quad
 \hat{S}_{l}^{-}=\hat{a}_{l}\prod_{j=1}^{l-1}(1-2\hat{n}_{j}),
\end{split}
\end{align}
maps the fermionic model Eq.~\eqref{eq:HamilrandomFermi} 
to a spin 1/2 chain with anisotropy along the $z$ axis, 
which is called XXZ model, in a random magnetic field. 
\begin{align} \label{eq:spinham}
 \hat{\mathcal{H}}_{\mathrm{XXZ}}
   =&J\sum_{l=1}^{N-1}
     (\hat{S}_{l}^{x}\hat{S}_{l+1}^{x}
     +\hat{S}_{l}^{y}\hat{S}_{l+1}^{y}
     +\Delta \hat{S}_{l}^{z}\hat{S}_{l+1}^{z})\nonumber\\
   &-\sum_{l=1}^{N}h_{l}\hat{S}_{l}^{z},
\end{align}
where $\hat{S}_{l}^{x,y,z}$ is the spin-1/2 operator 
and $S_{l}^{\pm}=S_{l}^{x}\pm i S_{l}^{y}$. 
For simplicity, we employ the unit $\hbar=c=1$. 
Additionally the spins can be mapped onto hard core bosons by the mapping $\hat{S}_{l}^{+}=\hat{b}_{l}^{\dagger}$ and $\hat{S}_{l}^{z}=\hat{b}_{l}^{\dagger}\hat{b}_{l}-\frac{1}{2}$. 

For $\Delta=0$, the Hamiltonian Eq.~\eqref{eq:HamilrandomFermi} 
represent free fermions with a random chemical potential. 
It is known that in one dimension, 
such a system is always exponentially localized for $W>0$~\cite{Anderson1958PR}.
The situation becomes complicated in the interacting case $\Delta\neq 0$,
but an analysis using field theory and bosonization
is possible~\cite{GiamarchiSchulz1988PRB,Giamarchi2004Book}. Such an analysis, using a RG procedure 
shows the existence of a quantum phase transition between a localized and delocalized phases  
The system is localized when $K$ is smaller than $3/2$ even for an infinitesimal $W$. 
The dependency of $K$ on $\Delta$ [Eq.~\eqref{eq:TLLparameter}] is shown in Fig.~\ref{fig:DelvsK} and it can be seen that 
$K<3/2$ corresponds to $\Delta>-1/2$. 

Besides the phase diagram itself the main physical quantity we will be computing in this paper is the frequency dependence of the conductivity. All calculations will be done at zero temperature. We use the Kubo formula relating the conductivity to current-current correlations. The current operator is
\begin{align}
 \hat{j}_{l}
   =\frac{iJ}{2}
     (\hat{a}_{l}^{\dagger}\hat{a}_{l+1}
     -\hat{a}_{l+1}^{\dagger}\hat{a}_{l}),
\end{align}
and the current-current retarded correlation function 
is written as
\begin{align}
 C_{\mathrm{cc}}(t)=-i\vartheta(t)
     \braket{[\hat{j}_{\mathrm{tot}}(t),\hat{j}_{\mathrm{tot}}(0)]},
\label{eq:RetCorr}
\end{align}
where $\vartheta(t)$ is the step function
and $\hat{j}_{\mathrm{tot}}=\sum_{l=1}^{N-1}\hat{j}_{l}$.
We denote by $\hat O(t)$ the usual Heisenberg time evolution 
$\hat O(t) = e^{i \hat{\mathcal{H}} t} \hat O e^{-i \hat{\mathcal{H}} t} $

The dynamical conductivity can be obtained 
by the Fourier transform
of the retarded correlation function Eq.~\eqref{eq:RetCorr},
\begin{align}
 \sigma(\omega)+i\sigma'(\omega)
   =\frac{i}{N\omega}\int_{-\infty}^{\infty}dt
     e^{i(\omega + i \epsilon) t}C_{\mathrm{cc}}(t),
\label{eq:OptConduct}
\end{align}
where $\sigma(\omega)$ and $\sigma'(\omega)$ represent 
the real and imaginary part 
of the dynamical conductivity, respectively. $\epsilon = 0^+$ is an infinitesimal convergence factor. 
Note that in the above expression we have not written the so-called diamagnetic term, which is purely imaginary, 
since we will concentrate on the real part of the conductivity $\sigma(\omega)$. 

In the spectral representation,
the real part of the conductivity is rewritten 
from Eq.~\eqref{eq:OptConduct} as
\begin{align}
 \sigma(\omega)
   =\frac{\pi}{N\omega}\sum_{\nu>0}
     \delta(\omega-E_{\nu}+E_{0})
     \big|\!\braket{\Psi_{\nu}|\hat{j}_{\mathrm{tot}}|\Psi_{0}}
     \!\big|^{2},
\label{eq:ConductLehmann}
\end{align}
where $\ket{\Psi_{0}}$ ($\ket{\Psi_{\nu}}$) 
is the ground ($\nu$-th excited) state 
and $E_{0}$ ($E_{\nu}$) is its energy eigenvalue. 
We will use the expression Eq.~\eqref{eq:ConductLehmann} 
for the numerical evaluation of the conductivity. 

\section{Methods}
\label{sec:methods}

We explain the numerical method that we use to treat
the interacting systems with disorder. 
In Sec.~\ref{sec:results}, we discuss the dynamical conductivity of such systems by comparing the results from this numerical method with
those from a field theory for the low energy limit of this model. 
The details of the field theory are described in Appendix~\ref{ap:bosonization}. 

To tackle the problem of low-dimensional interacting quantum systems, 
numerical methods utilizing matrix product states, 
such as DMRG~\cite{White1992PRL,Schollwock2011AnnPhys}, are very effective. 
The dynamical quantities such as conductivity and Green's function 
can be computed by performing a real-time evolution with e.g., time-evolving block decimation, after obtaining the ground state by DMRG, and such techniques have been widely used~\cite{CazalillaMarston2002PRL,Vidal_tebd2004PRL,Hallberg2006AdvPhys}. 
The spectral functions are calculable through the Fourier transformation of the temporal correlation functions.
However, in the real-time evolution of matrix product states, the entanglement of the systems grows exponentially and 
the achievable time interval is limited. The acquired frequency resolution in this way 
was not sufficient for our purpose.
Therefore we used a numerical method which calculate the spectral functions directly in the frequency space. 

In particular, to perform our calculation of the conductivity, 
we focus on the method CheMPS~\cite{Holzner2011PRB}.
This is a combination of DMRG and the kernel polynomial method~\cite{Weisse2006RMP,Holzner2011PRB},  a method to evaluate the spectral function 
\begin{align}
 A_{\hat{\mathcal{O}}_{1}\hat{\mathcal{O}}_{2}}(\omega)
   =\braket{\Psi_{0}|\hat{\mathcal{O}}_{1}
     \delta(\omega-\hat{\mathcal{H}}+E_{0})
     \hat{\mathcal{O}}_{2}|\Psi_{0}},
\label{eq:SpecFunc}
\end{align}
where $|\Psi_{0}\rangle$ is the ground state 
and $E_{0}$ is its energy. 
We assume that the spectra have nonzero weight in $\omega \in [0,\Omega]$,
which can be mapped to the interval
$\omega^{\prime}\in[-1+\epsilon_{\mathrm{s}},1-\epsilon_{\mathrm{s}}]$ 
by redefining the energy scale as 
\begin{align}
 \omega^{\prime}
   =\frac{2(1-\epsilon_{\mathrm{s}})}{\Omega}
     \omega-(1-\epsilon_{\mathrm{s}}),
\end{align}
where $\epsilon_{\mathrm{s}}$ is a small safety factor. 
We take $\epsilon_{\mathrm{s}}=0.0125$ in this study. 

The Hamiltonian is then mapped to
\begin{align}
 \hat{\mathcal{H}}^{\prime}
   =\frac{2(1-\epsilon_{\mathrm{s}})}{\Omega}
     (\hat{\mathcal{H}}-E_{0})
     -(1-\epsilon_{\mathrm{s}}),
\end{align}
and the spectral function Eq.~\eqref{eq:SpecFunc} becomes 
\begin{align}
 A_{\hat{\mathcal{O}}_{1}\hat{\mathcal{O}}_{2}}(\omega)
   =\frac{2(1-\epsilon_{\mathrm{s}})}{\Omega}
     \braket{\Psi_{0}|\hat{\mathcal{O}}_{1}
     \delta(\omega'-\hat{\mathcal{H}}')
     \hat{\mathcal{O}}_{2}|\Psi_{0}}.
\end{align}
Using the Chebyshev polynomials
\begin{align}
 T_{n}(\omega^{\prime})=\cos(n\arccos \omega^{\prime}),
\end{align}
we can expand the spectral function as 
\begin{align}
 A_{\hat{\mathcal{O}}_{1}\hat{\mathcal{O}}_{2}}(\omega)
   =&\frac{2(1-\epsilon_{\mathrm{s}})}{\Omega}
     \frac{1}{\pi\sqrt{1-{\omega^{\prime}}^{2}}}
     \Big[\mu_{0}+2\sum_{n=1}^{\infty}\mu_{n}T_{n}(\omega')\Big].
\label{eq:ChebyExpand}
\end{align}
The Chebyshev moments are represented as 
$\mu_{n}=\braket{\Psi_{0}|\hat{\mathcal{O}}_{1}|t_{n}}$, 
where $|t_{n}\rangle=T_{n}(\mathcal{H}')\hat{\mathcal{O}}_{2}\ket{\Psi_{0}}$
are Chebyshev vectors. 
The recurrence equations 
\begin{align}
\begin{split}
 &\ket{t_{n}}
   =2\hat{\mathcal{H}}'\ket{t_{n-1}}-\ket{t_{n-2}},\\
 &\ket{t_{0}}
   =\hat{\mathcal{O}}_{2}\ket{\Psi_{0}},\quad
 \ket{t_{1}}
   =\hat{\mathcal{H}}'\ket{t_{0}}
\end{split}
\label{eq:CheRecur}
\end{align}
are useful to evaluate the coefficients $\mu_{n}$ numerically. 
In the numerical calculations,
the expansion of Eq.~\eqref{eq:ChebyExpand} is performed 
up to some finite order $M$, 
and we multiply the weight $\mu_{n}$ by the Jackson damping factor~\cite{Weisse2006RMP}
\begin{align}
 g_{n}=\frac{(M-n+1)\cos\frac{n\pi}{M+1}
   +\sin\frac{n\pi}{M+1}\cot\frac{\pi}{M+1}}{M+1}
\end{align}
to smoothen the spectrum. 
Therefore, the spectral function is numerically 
obtained as 
\begin{align}
 A_{\hat{\mathcal{O}}_{1}\hat{\mathcal{O}}_{2}}(\omega)
   \simeq&\frac{2(1-\epsilon_{\mathrm{s}})}{\Omega}
     \frac{1}{\pi\sqrt{1-{\omega^{\prime}}^{2}}}\nonumber\\
     &\;\;\times\Big[g_{0}\mu_{0}
     +2\sum_{n=1}^{M}g_{n}\mu_{n}T_{n}(\omega^{\prime})\Big].
\end{align}

In this study, we calculate the ground state $\ket{\Psi_{0}}$ 
using DMRG, then obtain the matrix product state representation 
of $\ket{t_{n}}$ by the recurrence equations Eq.~\eqref{eq:CheRecur}. 
The system size is $N=250$, the energy width is $\Omega=6$, 
the bond dimension of matrix product representation is
$M_{\mathrm{B}}=64$, and the order of expansion is $M=200$.

The dynamical conductivity is calculated from 
the current-current correlation function as 
given by Eq.~\eqref{eq:OptConduct}. 
However, in the low energy region $\omega/J \ll 1$, 
the $1/\omega$ factor in the right hand side of 
Eq.~\eqref{eq:OptConduct} enhances the numerical error. 
Hence, to avoid this problem we employ the polarization-current correlation function instead of the current-current one. 
The polarization operator is defined as 
\begin{align}
 \hat{P}=\sum_{l=1}^{N} l\hat{n}_{l},
\nonumber
\end{align}
and it is related to the current operator 
through the time derivative 
\begin{align}
 \frac{\partial\hat{P}}{\partial t}
   =-i\sum_{l=1}^{N}
     l[\hat{n}_{l},\hat{\mathcal{H}}]
   =\hat{j}_{\mathrm{tot}}.
\nonumber
\end{align}
The polarization-current correlation function becomes 
\begin{align}
 C_{\mathrm{pc}}(t)=-i\vartheta(t)
     \braket{[\hat{P}(t),\hat{j}_{\mathrm{tot}}(0)]},
\nonumber
\end{align}
and its time derivative is the current-current correlation function 
\begin{align}
 \frac{\partial C_{\mathrm{pc}}(t)}{\partial t}
   =-i\vartheta(t)
     \braket{[\frac{\partial\hat{P}(t)}{\partial t},
     \hat{j}_{\mathrm{tot}}(0)]}
   =C_{\mathrm{cc}}(t).
\nonumber
\end{align}
Performing the integral by part in Eq.~\eqref{eq:OptConduct}, we obtain
\begin{align}
 \sigma(\omega)+i\sigma'(\omega)
   =&\frac{i}{N\omega}\int_{-\infty}^{\infty}dt
     e^{i\omega t}
     \frac{\partial C_{\mathrm{pc}}(t)}{\partial t}
\nonumber\\
   =&\frac{1}{N}\int_{-\infty}^{\infty}dt
     e^{i\omega t}C_{\mathrm{pc}}(t).
\nonumber
\end{align}
Thus, the dynamical conductivity is represented as 
\begin{align}
 \sigma(\omega)=\frac{\pi}{N}
   \braket{\Psi_{0}|\hat{P}
     \delta(\omega-\hat{\mathcal{H}}+E_{0})
     \hat{j}_{\mathrm{tot}}|\Psi_{0}},
\nonumber
\end{align}
which can be evaluated by CheMPS with 
$\hat{\mathcal{O}}_{1}=\hat{P}$ and
$\hat{\mathcal{O}}_{2}=\hat{j}_{\mathrm{tot}}$
in Eq.~\eqref{eq:SpecFunc}. 

\section{Results} \label{sec:results}

We now examine the results obtained by the numerical method described in Sec.~\ref{sec:methods}. 
Before dealing with the interacting case, which is the focus of our study, let us first discuss the dynamical conductivity in the noninteracting case, 
which corresponds to Anderson localization, 
where both simpler numerical solutions and 
analytical approaches are available.
Through the comparison of the results from CheMPS with from simpler methods,
we can provide a benchmark for the reliability and applicability of CheMPS. 

\subsection{Noninteracting case} \label{sec:noninteracting}

Let us first revisit the noninteracting case $\Delta=0$. 
Without interactions, Eq.~\eqref{eq:HamilrandomFermi} 
is a tight binding system of free spinless fermions at half-filling.

If one linearizes the dispersion relation 
around the Fermi wave number $k=\pm \pi/2$
in the continuum limit, 
this model reduces to a Dirac model with a random chemical potential 
for which an exact analytical solution was obtained 
by Berezinskii~\cite{Berezinskii1974JTEP} 
and the conductivity could be computed analytically~\cite{Berezinskii1974JTEP,AbrikosovRyzhkin1978AdvPhys,Gogolin1978PSS,Gogolin1982PhysRep}.  
 
The expression for the dynamical conductivity is given as~\cite{Gogolin1982PhysRep} 
\begin{align}
 \sigma(\omega)=\sigma_{0}\sum_{n=0}^{\infty}
   Q_{n}(R_{n}-R_{n+1}),
\label{eq:GogolinFormula}
\end{align}
where $Q_{n}$ and $R_{n}$ are the solution of 
the recurrence equations
\begin{align}
 &2i\omega\tau_{\mathrm{i}}R_{n}+n(R_{n+1}+R_{n-1}-2R_{n})=0
\label{eq:GogolinRecur1}\\
 &2i\omega\tau_{\mathrm{i}}(n+1/2)Q_{n}+(n+1)^{2}(Q_{n+1}-Q_{n})
\nonumber\\
   &\qquad-n^{2}(Q_{n}-Q_{n-1})+R_{n}-R_{n+1}=0
\label{eq:GogolinRecur2}
\end{align}
with the boundary condition
\begin{align}
 R_{0}=1,\quad
 i\omega\tau_{\mathrm{i}}Q_{0}+Q_{1}-Q_{0}+R_{0}-R_{1}=0
\label{eq:GogolinBound}
\end{align}
and 
\begin{align}
 \lim_{n\to\infty}Q_{n}=0,\quad
 \lim_{n\to\infty}R_{n}=0.
\nonumber
\end{align}

\begin{figure}[t]
\centering
\includegraphics[width=0.3\textwidth]{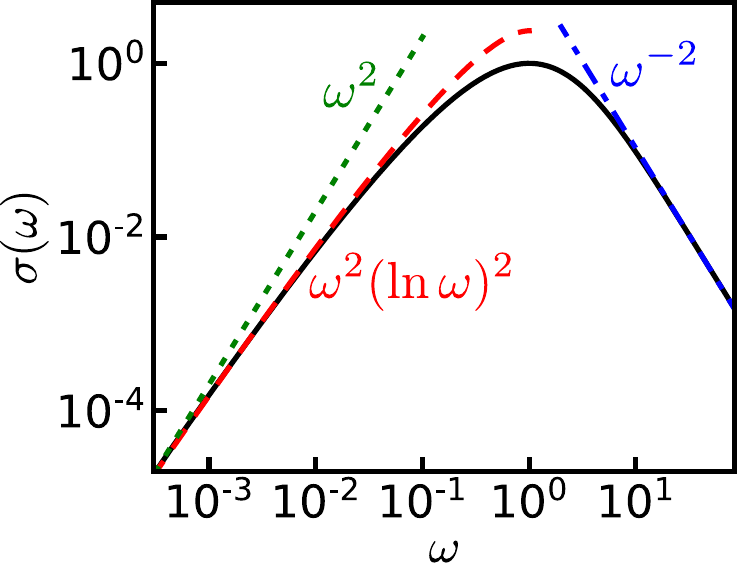}
\caption{\label{fig:Berezinskii}
Dynamical conductivity calculated from 
Eq.~\eqref{eq:GogolinFormula} (solid line). 
The spectrum is scaled so that the maximum of $\sigma(\omega)$ and its frequency become $\sigma(\omega)=1$ and $\omega=1$. 
The fitting curves $\sigma(\omega)=17.5(\omega/e)^{2}(\ln(\omega/e))^{2}$ (dashed line), $\sigma(\omega)=200\omega^{2}$ (dotted line), 
and $\sigma(\omega)=10.5\omega^{-2}$ (dashed-dotted line) are shown. 
The function $\omega^{2}(\ln\omega)^{2}$ is also scaled so that 
the peak is located at $\omega=1$ by $\omega\to \omega/e$.}
\end{figure}

In practice, starting from the initial condition 
$Q_{n}=0$ and $R_{n}=0$ for large enough $n$, 
we can obtain numerically 
$Q_{n-1},\ldots,Q_{0}$ and $R_{n-1},\ldots,R_{0}$ 
from Eqs.~\eqref{eq:GogolinRecur1} and \eqref{eq:GogolinRecur2}, 
and finally normalize the sequence 
$\{Q_{0},\ldots,Q_{n}\}$ and $\{R_{0},\ldots,R_{n}\}$ 
so as to satisfy the condition Eq.~\eqref{eq:GogolinBound}.
Here, $\tau_{\mathrm{i}}$ and $\sigma_{0}$
are fitting parameters.
Fig.~\ref{fig:Berezinskii} shows the dynamical conductivity $\sigma(\omega)$ calculated by the above procedure. 
In the high frequency region, the dynamical conductivity decays as a power law $\sigma(\omega)\propto \omega^{-2}$. 
In the low frequency region, the predicted analytical behavior is $\sigma(\omega)\propto \omega^{2}(\ln\omega)^{2}$~\cite{Mott1970PhilMag,Berezinskii1974JTEP,AbrikosovRyzhkin1978AdvPhys}.
As can be seen from Fig.~\ref{fig:Berezinskii}, such a behavior fits much better the data than 
$\sigma(\omega)\propto \omega^{2}$. 

%%%%%%%%%% Fig : Berezinskii Solution %%%%%%%%%%
%%%%%%%%%%%%%%%%%%%%

\begin{figure}[t]
\includegraphics[width=0.48\textwidth]{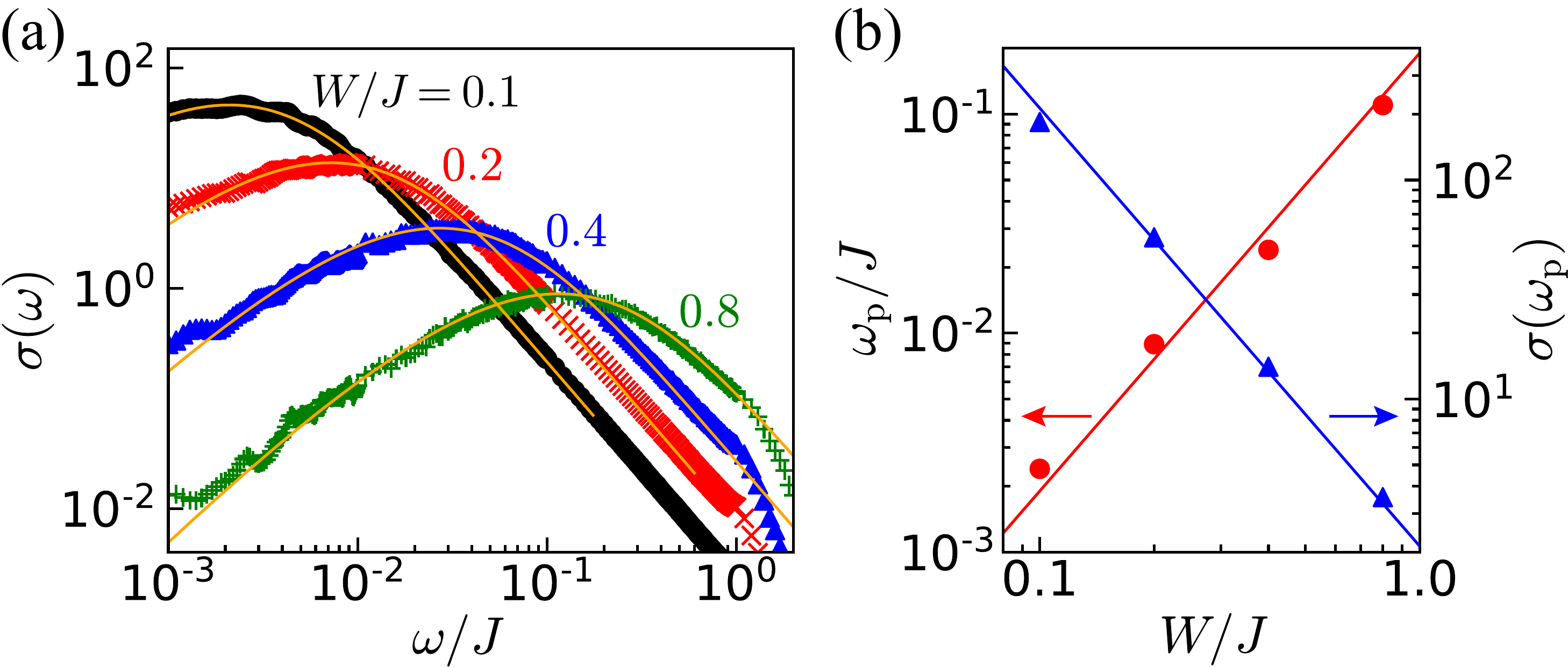}
\caption{
(a) Dynamical conductivity for the noninteracting case 
$\Delta=0$ as a function of the frequency $\omega$ for strengths of the disorder 
$W/J=0.1$, $0.2$, $0.4$, and $0.8$ calculated by ED (see text). 
The solid lines are fitting curves calculated 
from Eq.~\eqref{eq:GogolinFormula} the continuum (Dirac) model. 
(b) The pinning frequency $\omega_{\mathrm{p}}$ and 
the corresponding value of the conductivity at this frequency 
$\sigma(\omega_{\mathrm{p}})$ obtained from the curves 
shown in the panel (a). 
The solid lines represent the fitting curves 
$\omega_{\mathrm{p}}/J=0.19(W/J)^{2}$ and 
$\sigma(\omega_{\mathrm{p}})=2.13(W/J)^{-2}$.}
\label{fig:freeED}
\end{figure}

To compare with the CheMPS calculations, 
we also perform exact diagonalization directly 
on the lattice model Eq.~\eqref{eq:HamilrandomFermi} 
with $\Delta=0$. We evaluate the dynamical conductivity by 
\begin{align}
 \sigma(\omega)
   =\frac{1}{N\omega}\sum_{\nu>0}
     \frac{\delta_{0}}{(\omega-E_{\nu}+E_{0})^{2}+\delta_{0}^{2}}
     \big|
     \braket{\Psi_{\nu}|\hat{j}_{\mathrm{tot}}|\Psi_{0}}
     \big|^{2}.
\nonumber
\end{align}
In this equation, we have introduced 
a nonzero broadening $\delta_{0}$  to the delta function 
in Eq.~\eqref{eq:ConductLehmann}. 
The size of the noninteracting system is $N=3200$, 
the broadening is $\delta_{0}/J=2\times 10^{-4}$, 
and we take the ensemble average of calculations 
for $800-1600$ configurations of the random chemical potential. 
The dynamical conductivities for disorder strengths 
$W/J=0.1$, $0.2$, $0.4$, and $0.8$ are shown in Fig.~\ref{fig:freeED}(a). 
The data for each value of $W$ is fitted with the formula of the linearized dispersion model Eq.~\eqref{eq:GogolinFormula} 
with the fitting parameters $\tau_{\mathrm{i}}$ and $\sigma_{0}$. 

The dynamical conductivity of the free fermion model on the lattice 
is well fitted by the curve obtained 
from Eq.~\eqref{eq:GogolinFormula}, 
confirming that the lattice model correctly captures the 
$\sigma(\omega)\propto \omega^{2}(\ln\omega)^{2}$ behavior
in the low frequency region and $\sigma(\omega)\propto \omega^{-2}$ 
in the high frequency region. 
Given the finite bandwidth $J$ of the lattice model, 
above $\omega/J>1$, $\sigma(\omega)$ decays exponentially 
and deviates from the curve obtained from Eq.~\eqref{eq:GogolinFormula}. 

In addition to the asymptotic behaviors at small and large frequency we can extract the pinning frequency 
$\omega_{\mathrm{p}}$ and the peak value $\sigma(\omega_{\mathrm{p}})$ 
from the dynamical conductivity curves shown in Fig.~\ref{fig:freeED}(a). 
The quantities $\omega_{\mathrm{p}}$ and $\sigma(\omega_{\mathrm{p}})$ are plotted in Fig.~\ref{fig:freeED}(b) as a function of $W$. 
We observe that 
$\omega_{\mathrm{p}}$ and $\sigma(\omega_{\mathrm{p}})$ are scaled with the disorder strength $W$ as 
\begin{align}
 \omega_{\mathrm{p}}\propto W^{2},\quad
 \sigma(\omega_{\mathrm{p}})\propto W^{-2}.
 \label{eq:peakscaleW}
\end{align}
This is the expected behavior since the pinning frequency 
is directly related to the inverse localization length, 
which for the noninteracting case scales 
as the mean free path in one dimension 
and thus as $\xi\propto W^{-2}$, 
as can be seen from Eqs.~\eqref{eq:peakscale} and \eqref{eq:peakscaleW}.

%%%%%%%%%% Fig : Dynamical conductivity for free fermions %%%%%%%%%%
%%%%%%%%%%%%%%%%%%%%

The results for the noninteracting case also serve as a benchmark of the CheMPS method in the next section. 

\subsection{Interacting case}
\label{sec:interacting}

Let us now turn to the interacting disordered systems. In this case, since the dimension of the Hilbert space grows exponentially by increasing the system size, ED is not a viable method any more and we employ the CheMPS method described in the Sec.~\ref{sec:methods} as a numerical approach as well as the bosonization and variational replica approach as analytical methods to calculate the dynamical conductivity. 

%%%%%%%%%% Fig : Comparison between CheMPS and ED %%%%%%%%%%
\begin{figure}[t]
\includegraphics[width=0.48\textwidth]{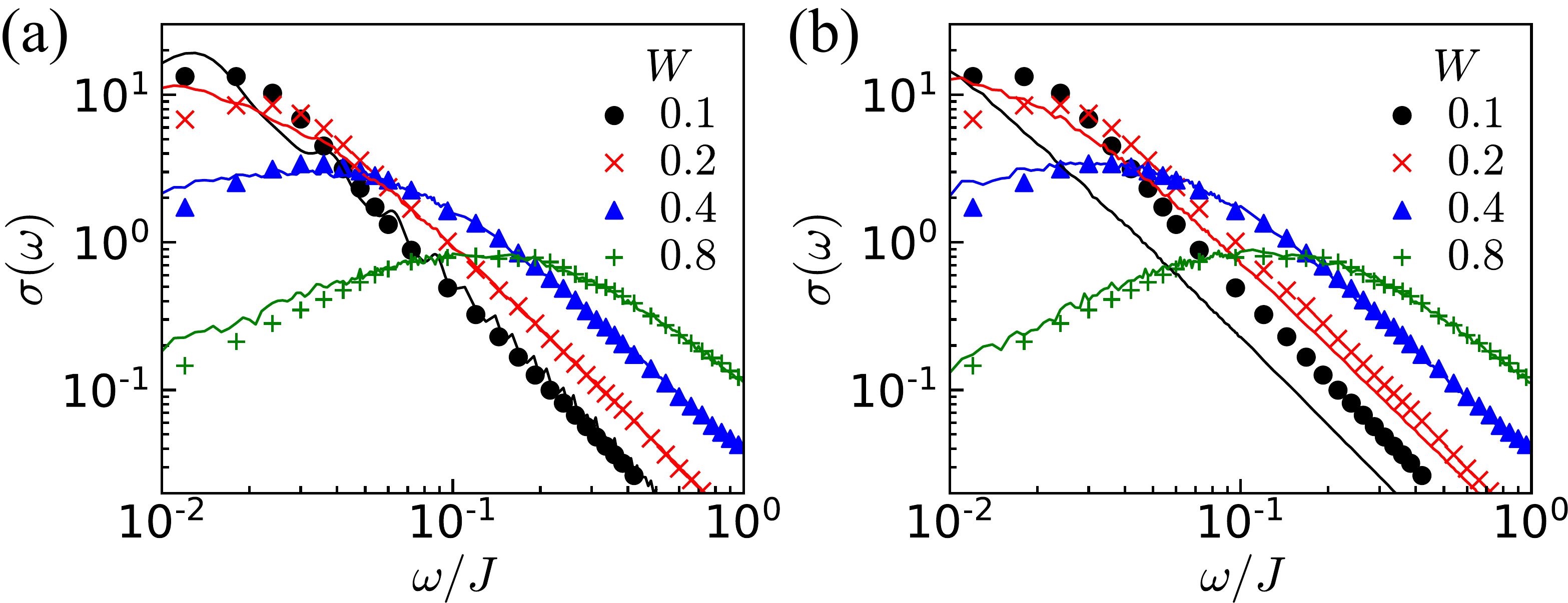}
\caption{Comparison between CheMPS (marks) 
with $N=250$ and ED (solid lines) 
with (a) $N=250$ and (b) $N=3200$ 
for the dynamical conductivity in the noninteracting case 
$\Delta=0$ with disorders $W/J=0.1$, $0.2$, $0.4$, and $0.8$}
\label{fig:compare}
\end{figure}
%%%%%%%%%%%%%%%%%%%%

To benchmark the method, let us first compare the results of CheMPS for system size $N=250$ with $\Delta=0$ and disorders $W/J=0.1$, $0.2$, $0.4$, and $0.8$ 
with the ones obtained in the previous section with ED. 
The comparison is shown in Fig.~\ref{fig:compare}. 
As can be seen from the comparison, the agreement is good over two decades of frequency, particularly in the regime $\omega/J>0.05$. 
To check the finite size effect, we also compare the same CheMPS data 
with the ED result for $N=3200$  (the same data as Fig.~\ref{fig:freeED}) 
in Fig.~\ref{fig:compare}(b). While the agreement is good for large $W$, 
there is a deviation for small $W$. However, in the high energy region, 
the deviation is just a multiplication of a constant factor, 
and the power of the decay does not change. 
This confirms that our numerical approach properly captures the behavior of the dynamical conductivity in both low and high frequency regions at zero temperature. 
Hence CheMPS is a very useful technique for dealing with interacting systems, 
for which no other quantitative method is available. 

\begin{figure}[t]
\includegraphics[width=0.48\textwidth]{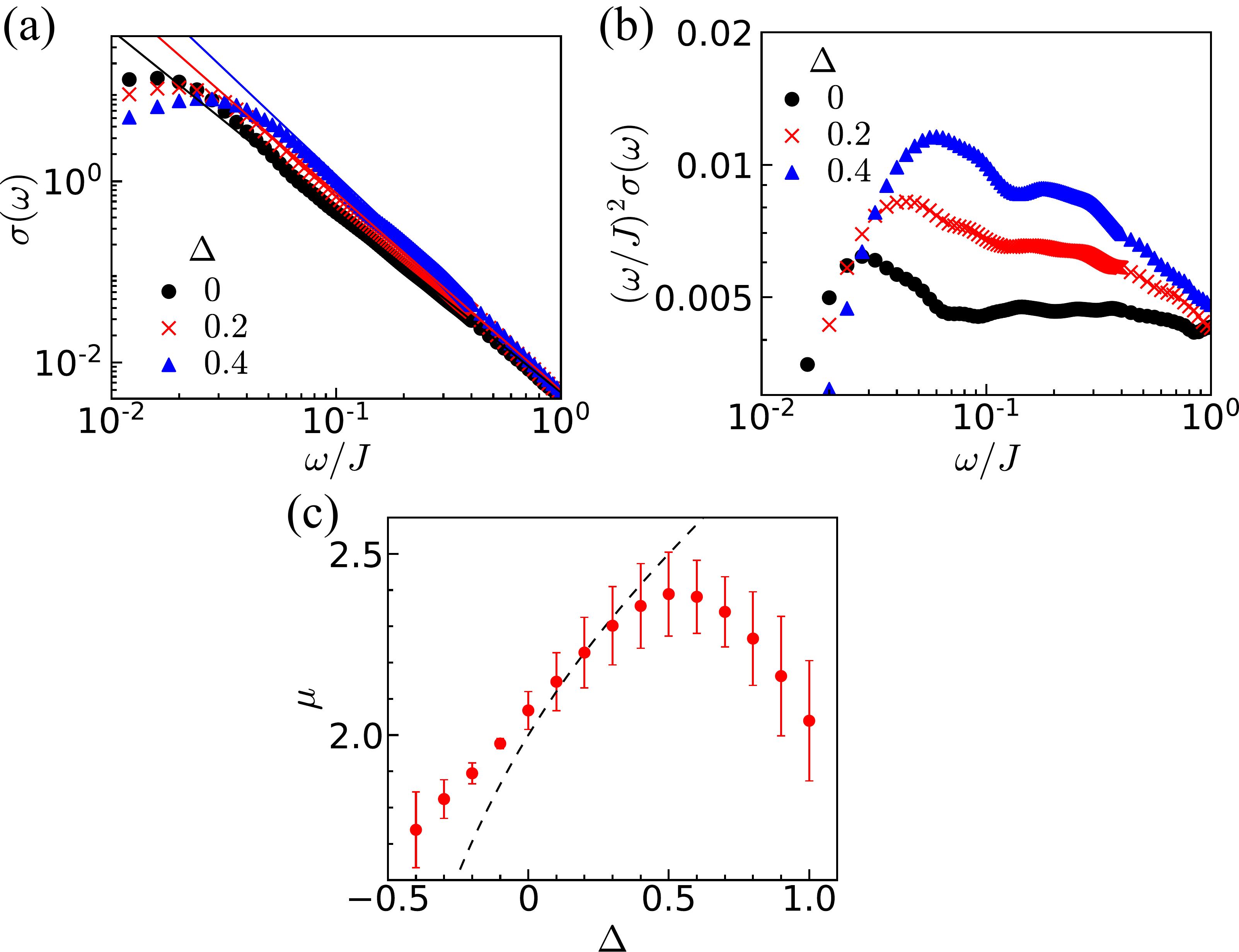}
\caption{\label{fig:Power_HighFreq}
(a) The dynamical conductivity $\sigma(\omega)$ 
for interactions $\Delta=0$, 0.2, and 0.4, 
and the disorder strength $W/J=0.1$. 
The solid lines represent the power-law fitting 
in the high energy region. The frequency dependence of the conductivity is well represented by an interaction dependent exponent at high frequencies. 
(b) The plotting of $(\omega/J)^{2}\sigma(\omega)$ for the same data as Fig.~\ref{fig:Power_HighFreq}(a).
(c) Exponent, as a function of the interaction $\Delta$, of the decay of the conductivity with frequency in the high frequency region.
This exponent results from fits $\sigma(\omega)\propto \omega^{-\mu}$ similar to the ones of (a). The disorder strength is $W/J=0.1$.
The dashed line is the theoretical value of the exponent from the field theory $\mu=4-2K$ and the Bethe-ansatz value of the TLL parameter $K$ (see text). 
For repulsive interactions $\Delta > 0$ there is a reasonable agreement up to $\Delta \sim 0.5$ beyond which there is a plateau like behavior 
not expected by the field theory. On the attractive side $\Delta < 0$ strong deviations are observed even for relatively small attraction.}
\end{figure}

In Fig.~\ref{fig:Power_HighFreq}(a), we show the numerically calculated dynamical conductivity for the cases of $\Delta=0$, 0.2, and 0.4. We can see that the decay power of the high frequency region changes as we increase the interaction $\Delta$. To discuss the high frequency region and to compare the results with the bosonized field theory, a relatively small disorder strength is desirable, and we adopt $W/J=0.1$ here. 
We fit the data in the high frequency region by a power law $\propto \omega^{-\mu}$ and plot $\mu$ as a function of $\Delta$ in Fig.~\ref{fig:Power_HighFreq}(c). 
We confirm that the variance of the data for dynamical conductivity shown in Fig.~\ref{fig:Power_HighFreq}(a) is negligibly small by comparing the results for three bins of the various disorder configurations over which the average is taken. The variations of the data in Fig.~\ref{fig:Power_HighFreq}(c) mainly arise from the power law fitting, and we estimate the error bars from the fittings in several frequency intervals on the high frequency regime.

Let us compare the numerical results with the prediction 
from the field theory on the continuum model~\cite{GiamarchiSchulz1988PRB} (see Appendix~\ref{ap:bosonization}). 
The behavior in the high frequency regime remains a power law while the exponent is indeed modified by introducing the interaction. 
Note that the precise analytical form is more complicated than a simple power law since the TLL parameter $K$ is renormalized and depends on the scale. However, far from the transition point $K=3/2$, the simple power law decay which neglects this renormalization becomes a good approximation~\cite{GiamarchiSchulz1988PRB,Giamarchi2004Book} .
The modification of the exponent by the interaction reflects the renormalization of the scattering on the disorder and the power law behavior of the susceptibility of the charge-charge correlations in TLL. 
The analytical prediction for the exponent is $\mu = 4 - 2K$~\cite{GiamarchiSchulz1988PRB,Giamarchi2004Book}, which naturally reproduces the exponent $\mu = 2$ for the noninteracting case $K=1$. The parameter $K$ takes the value of $K<1$ for repulsive interactions and $K>1$ for attractive ones. 
As seen in Fig.~\ref{fig:Power_HighFreq}(c), 
the numerically evaluated $\mu$ has a similar global trend as the expectations from the field theory (dashed line), which is obtained by substituting the Bethe ansatz evaluation of $K$ for TLL in the XXZ chain Eq.~\eqref{eq:TLLparameter} into $\mu = 4 - 2K$. 
In particular, in the region of $-0.2\lesssim\Delta\lesssim 0.5$, the numerically calculated $\mu$ agrees quantitatively well with the analytical prediction. 
Figure~\ref{fig:Power_HighFreq}(b) shows the plotting of $(\omega/J)^{2}\sigma(\omega)$ for the same data as Fig.~\ref{fig:Power_HighFreq}(a). The plotting of the data for $\Delta=0$ is almost horizontal, which indicates the decay follows $\sigma(\omega)\propto \omega^{-2}$. We can see that the decay power $\mu$ increases as $\Delta$ becomes larger.

%%%%%%%%%% Fig : Power %%%%%%%%%%
%%%%%%%%%%%%%%%%%%%%

However, there exist surprising quantitative differences between the numerical 
results and the field theoretical predictions in the attractive regime 
($\Delta\lesssim-0.2$) and the strong repulsive regime ($\Delta\gtrsim 0.5$).
The origin of these discrepancies is not clear at the present stage.
On the repulsive side, a plateau-like structure appears in the regime 
$\Delta\gtrsim 0.5$ for the numerical results, which is incompatible with the exponent predicted by the continuum model.
Several reasons are conceivable for this behavior. 
One possibility is the effect of irrelevant operators neglected in the field theoretical treatment. In particular, the scaling dimension of the irrelevant operator $\cos(4\phi)$ representing
the umklapp processes on the lattice lowers as $\Delta$ is increased, and it finally becomes marginal in the $\Delta\to 1$ limit. 
Another possibility is an error of the numerical extraction of $\mu$. 
As seen from Fig.~\ref{fig:Power_HighFreq}(a), the localization frequency scale, i.e., the pinning frequency characterized by the peak of the dynamical conductivity, shifts to higher frequency as $\Delta$ is increased. Hence the region where the curve is fitted by the power law becomes narrower, and the estimation becomes less precise.
On the attractive side, if we approach the localization-delocalization transition point $K=3/2$ ($\Delta=-0.5$), the renormalization flow is closer to the separatrix and the direction of the flow is not parallel to the disorder axis. Thus the effect of renormalization of the parameter $K$ should become more important. To elucidate the reasons for this discrepancy between the numerics and the field theory is a very challenging problem and we leave it for a future study. 

Let us now turn to the disorder dependence of the pinning frequency and the conductivity at the peak for the interacting system. 
In Fig.~\ref{fig:sigma_Del0_5}(a), we show the dynamical conductivity calculated for the interaction $\Delta=0.5$ and various disorder strengths $W/J=0.1$, $0.2$, $0.4$, $0.8$. 
%%%%%%%%%% Fig : Dynamical conductivity for Delta=0.5 %%%%%%%%%%
\begin{figure}[t]
\includegraphics[width=0.48\textwidth]{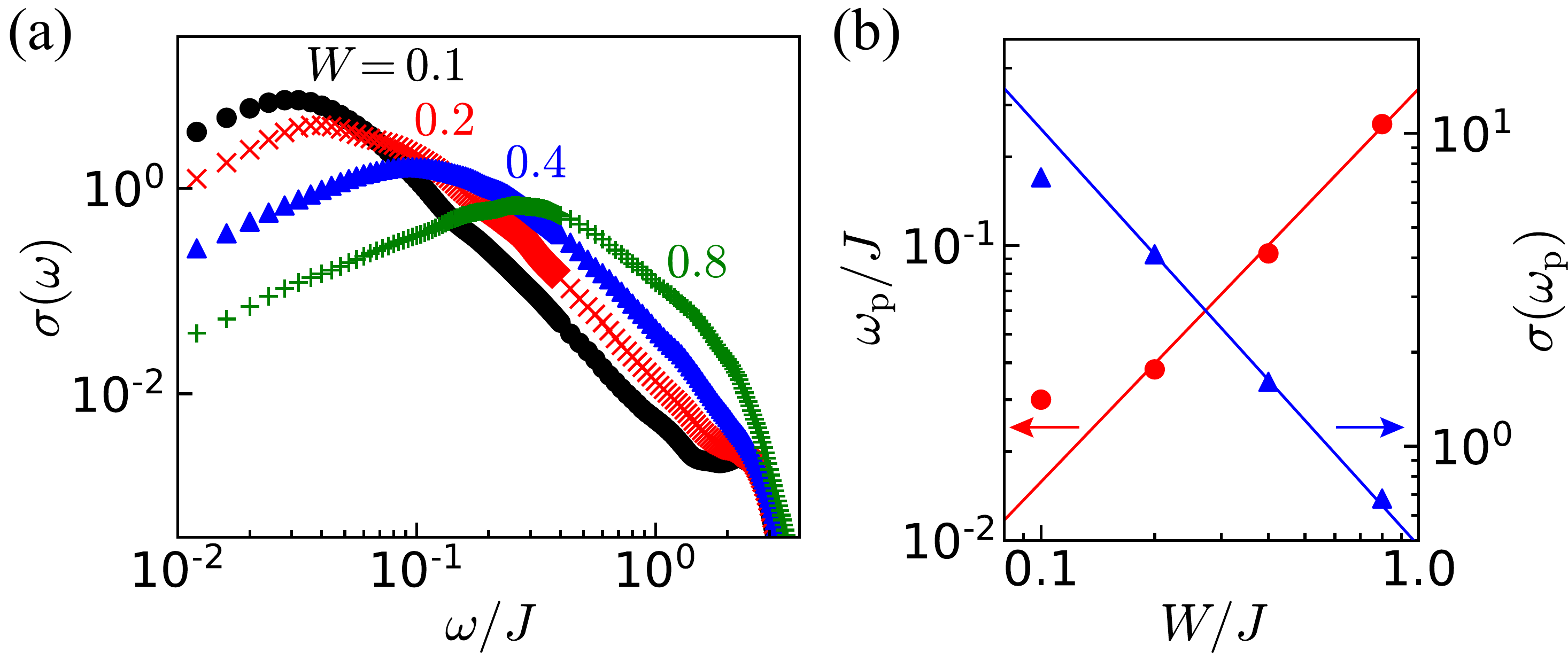}
\caption{(a) Dynamical conductivity calculated for the interaction $\Delta=0.5$  with disorder strengths $W/J=0.1$, $0.2$, $0.4$, and $0.8$. as for the noninteracting case one
observes a peak of the conductivity at a pinning frequency $\omega_{\mathrm p}$. 
(b) The pinning frequency $\omega_{\mathrm p}$ and 
the corresponding value of the conductivity at this frequency 
$\sigma(\omega_{\mathrm p})$ obtained from the data in the panel (a).  
The solid lines represent the fitting curves 
$\omega_{\mathrm{p}}/J=0.34(W/J)^{4/3}$ and 
$\sigma(\omega_{\mathrm{p}})=0.48(W/J)^{-4/3}$.}
\label{fig:sigma_Del0_5}
\end{figure}
%%%%%%%%%%%%%%%%%%%%
Similarly than for the noninteracting case [see Fig.~\ref{fig:freeED}(a)],  
the pinning frequency $\omega_{\mathrm{p}}$ increases and the peak height $\sigma(\omega_{\mathrm{p}})$ decreases as the disorder strength $W$ is increased. 
We plot $\omega_{\mathrm{p}}$ and $\sigma(\omega_{\mathrm{p}})$ as a function of $W$ in Fig.~\ref{fig:freeED}(b). 
The pinning frequency and the peak height are well fitted as 
\begin{align}
 \omega_{\mathrm{p}}\propto W^{4/3},\quad
 \sigma(\omega_{\mathrm{p}})\propto W^{-4/3}.
\end{align}
Note that the data points of $\omega_{\mathrm{p}}$ and $\sigma(\omega_{\mathrm{p}})$ for $W/J=0.1$ deviate from the fitting curves. We attribute it to the large finite size effect in the case of small $W$, as mentioned in the benchmark result [Fig.~\ref{fig:compare}(b)].

This scaling is in good agreement with the analytical predictions that $\omega_{\mathrm{p}}\propto\xi^{-1}$ and $\sigma(\omega_{\mathrm{p}})\propto\xi$ 
[See Eq.~\eqref{eq:peakscale} in Appendix~\ref{ap:bosonization}].
Using the dependence of $K$ on interactions Eq.~\eqref{eq:TLLparameter}, 
we obtain $K=3/4$ for $\Delta=0.5$.
This leads to $\xi\propto W^{-4/3}$ 
by using the formula Eq.~\eqref{eq:locdep} 
in excellent agreement with the numerical results. 
As can be seen both from the numerics and the above formula, 
repulsive interactions lead to a shorter localization length 
than for the noninteracting case. 

Finally let us discuss the behavior in the low frequency regime. 
In order to get a sizable range of the frequency interval below the pinning peak, and to prevent finite size effects from playing a major role, 
we take a relatively large value of disorder $W/J=0.8$. 
The dynamical conductivity calculated for $\Delta=0$, $0.3$, $0.5$, and $0.8$ is shown in Fig.~\ref{fig:sigma_Del0_5}(a). 
While there is a clear dependence of the exponent of the power law decay on the interaction $\Delta$ in the high frequency regime, the behavior remains similar on the low frequency side. One can see it more clearly in Fig.~\ref{fig:sigma_Del0_5}(b), where the curves have been rescaled by the value of the pinning frequency and conductivity at the peak.
%%%%%%%%%% Fig : Dynamical conductivity for W=0.8 %%%%%%%%%%
\begin{figure}[t]
\includegraphics[width=0.48\textwidth]{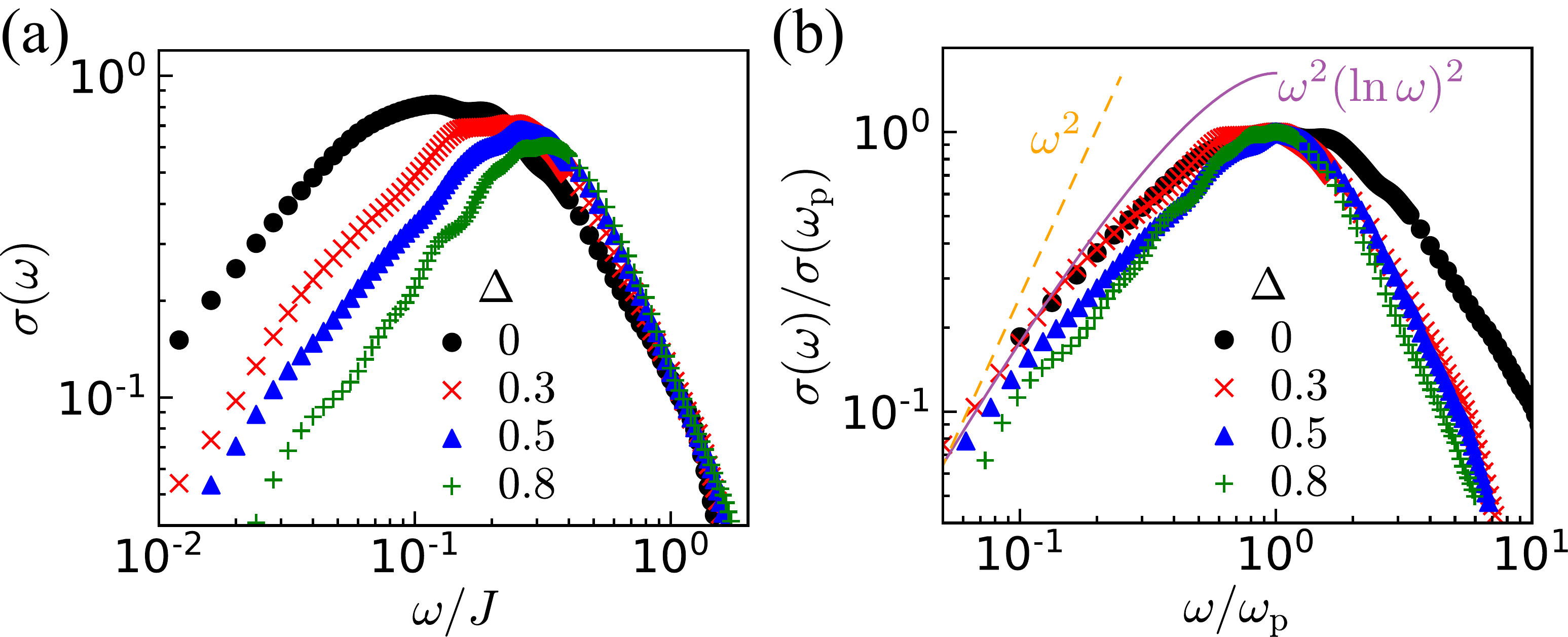}
\caption{\label{fig:sigma_Del0_5}
(a) Dynamical conductivity calculated 
for the disorder strength $W/J=0.8$ 
with the interaction $\Delta=0$, 0.3, 0.5, and 0.8.
(b) Dynamical conductivity normalized by the value 
at the pinning frequency $\sigma(\omega)/\sigma(\omega_{\mathrm{p}})$. 
The solid and dashed lines represent the curves 
$\propto \omega^{2}(\ln\omega)^{2}$ and $\propto \omega^{2}$, respectively.}
\end{figure}
%%%%%%%%%%%%%%%%%%%%

The fitting curves by $\sigma(\omega)\propto \omega^{2}(\ln\omega)^{2}$ 
and $\sigma(\omega)\propto \omega^{2}$ are also shown, and the former fitting looks better than the latter. 
However, the present interval of data fitting below the peak is just one decade of frequency, and the acquisition of the data over a wider range is desirable for a more precise analysis.

On a qualitative side one would indeed expect to recover at 
low enough frequencies essentially the noninteracting behavior. Indeed 
frequencies lower than the pinning frequency correspond to 
probing scales large compared to the localization length. 
At that scales since the particles are exponentially localized, 
the effect of interactions should drastically decrease, 
leading back to the noninteracting behavior. 
On a more quantitative level, it is difficult to make an unambiguous comparison with existing analytical formulas, since 
the various calculations of the low frequency behavior suffer from their own limitations. 
The variational calculation \cite{GiamarchiLeDoussal1996PRB} is unable to capture the logarithmic correction to the $\omega^{2}$ behavior. Calculations containing the logarithmic correction rely either on an extreme 
classical limit $K \to 0$~\cite{Fogler2002PRL} 
which is far from the values of $K$ reached here 
or an instanton expansion~\cite{Nattermann2003PRL,RosenowNattermann2006PRB}. 

Given the rescaling of the curves in Fig.~\ref{fig:sigma_Del0_5} and the 
fact that the different curves broadly superimpose, this suggests that 
the prefactor of the $\omega^{2}(\ln \omega)^{2}$ term varies as
\begin{align}
 \frac{\sigma(\omega_{\mathrm{p}})}{\omega_{\mathrm{p}}^{2}}
   \propto \bigg(\frac{\xi(\Delta)}{\xi(\Delta=0)}\bigg)^{3}
\end{align}
from the analytical predictions $\omega_{\mathrm{p}}\propto\xi^{-1}$ and $\sigma(\omega_{\mathrm{p}})\propto\xi$ 
[See Eq.~\eqref{eq:peakscale} in Appendix~\ref{ap:bosonization}].

\section{Discussions and summary}
\label{sec:summary}

In this paper, we have numerically computed the frequency dependence 
of the conductivity in one-dimensional spinless fermions 
with a nearest neighbor interaction and a random chemical potential. 
This problem is equivalent to XXZ spin chains in a random magnetic field 
along the $z$ axis and to hard core bosons with nearest neighbor interactions 
and a random chemical potential. 
Using the CheMPS method, a variant of DMRG, 
adapted to the case we study, 
we have numerically calculated the dynamical conductivity 
over a broad range of frequencies, interactions, 
and disorder strength. We have benchmarked our method by comparisons 
with the noninteracting case. 
Since analytical approaches and numerical exact diagonalizations 
are applicable for the noninteracting systems, 
these results have been compared with those from CheMPS 
and we have confirmed that our method does capture the frequency dependence 
of the conductivity in a broad range of frequencies. 

We have then investigated the dynamical conductivity of the interacting systems with CheMPS. In the high frequency regime, the conductivity follows a power law $\sigma(\omega)\propto \omega^{-\mu}$. 
We have calculated $\mu$ for various interaction $\Delta$, and found it agrees quantitatively with the expectation from the field theory $\mu=4-2K$ and the $K$ value from BA in $-0.2\lesssim\Delta\lesssim 0.5$. 
However there exist a deviation in the attractive and strongly repulsive regions. 
On the attractive side $\Delta\lesssim -0.2$ ($K\gtrsim 1.15$), 
the estimated exponent $\mu$ seems larger than expected from the TLL determination. On the repulsive side, a plateau-like structure appears in the region $\Delta\gtrsim 0.5$ ($0.5\leq K\lesssim 0.75$). 
We leave the clarification of the reasons for this deviation for future studies. 
We have also evaluated the localization length $\xi$ from the pinning frequency $\omega_{\mathrm{p}}\propto\xi^{-1}$ as a function of the disorder strength, and found a reasonable agreement with the expected behavior of the localization length as determined by RG: 
$\xi\propto (1/W^{2})^{1/(3-2K)}$ and which is now dependent on both interactions and disorder. 
In the low frequency regime, we have performed numerical calculations for a large disorder $W/J=0.8$. The scaling of dynamical conductivity is compatible with an $\omega^{2}(\ln\omega)^{2}$ behavior similar to the one of the noninteracting case but with a prefactor varying as $(\xi(\Delta)/\xi(\Delta=0))^{3}$.

The low frequency behavior is difficult to access and although the numerics is indeed compatible with the $\omega^{2}(\ln\omega)^{2}$
behavior to ascertain the existence and power of the log correction data over a wider range of frequency is needed, 
another challenge for future studies. 

In addition to pushing the numerical investigations of the conductivity, it would of course be extremely interesting 
to test the results obtained in our study in cold atom experiments. The existence of a peak as a response to shaking, 
similar to the pinning peak in the conductivity discussed here, was indeed observed for bosons in a quasiperiodic potential and has been used as a key signature of the existence of the Bose glass in these systems. However, due to the large inhomogeneities arising from the quadratic trapping potential, testing the power law behavior was not practically feasible. The existence of fermionic systems with disorder in quantum microscopes makes it possible to observe the features computed here more easily. In preforming a comparison with experiments, we need to note the following points: i) the experimental system size realizable for the moment are still relatively small, typically of the order of $20$ to $50$ atoms per chain; ii) the temperature is still relatively high in fermionic systems. 
These conditions should not drastically affect the properties in the 
high frequency regime, roughly for $\omega > T$,  but of course will 
essentially change the low frequency behavior of the dynamical 
conductivity. The low frequency regime is ultimately connected to the question of variable 
range hopping and many-body localization. Addressing these issues via experiments and by the extension of numerical techniques to finite temperature is clearly a considerable challenge. 

\acknowledgements

The authors appreciate fruitful discussions with Michele Filippone.
S. T. is supported by JSPS KAKENHI Grant No. JP21K03412 
and JST CREST Grant No. JPMJCR19T3, Japan. 
This work is partly supported by the Swiss National Science Foundation under Division II. 

This paper is dedicated to Alain Aspect, who in addition to his numerous contributions in atomic physics and quantum information is responsible, via key experiments, to having made, the subject of disorder a very hot topic in cold atomic systems. Thank you Alain for being a constant source of inspiration and stimulation, also in this field of ``dirty'' quantum systems.  

\appendix

\section{Field theory solution}
\label{ap:bosonization}

As a framework to discuss the numerical solution, let us give here a short summary of the field theory approach to this problem~\cite{GiamarchiSchulz1988PRB}. 
The most convenient way to deal with the interactions is to use the bosonization representation~\cite{Giamarchi2004Book}. 

Through the Fourier transformation of the fermionic operators 
(the lattice constant is set to unity):
\begin{align}
 \hat{a}_{k}=\frac{1}{\sqrt{N}}\sum_{l}e^{-ikl}\hat{a}_{l},\quad
 \hat{a}_{k}^{\dagger}
    =\frac{1}{\sqrt{N}}\sum_{l}e^{ikl}\hat{a}_{l}^{\dagger},
\end{align}
the disorder term in the Hamiltonian becomes 
\begin{align}
 \hat{\mathcal{H}}_{\mathrm{dis}}
    =-\sum_{l=1}^{N}h_{l}\Big(\hat{n}_{l}-\frac{1}{2}\Big)
    =-\frac{1}{N}\sum_{k,q}h(q)\hat{a}_{k+q}^{\dagger}\hat{a}_{k},
\end{align}
Fermionic systems Eq.~\eqref{eq:HamilrandomFermi} with a Fermi momentum $k_{\mathrm{F}} = \pi/2$ (half-filled band) corresponds to the nonmagnetized case in spin chains. 
For such systems special care should be exerted to treat the disorder and the situation is slightly more involved than starting from an incommensurate filling. 
As is well-known in one dimension, the system has low energy excitations at momenta $q\simeq 0$ and $q\simeq 2k_{\mathrm{F}}$~\cite{Giamarchi2004Book}. It is thus useful to separate the disorder into two slowly varying fields centered around these momenta, 
which are called forward and backward scattering respectively,
\begin{align}
 \begin{split}
    h_{\mathrm{f}}(x) &= \sum_{q \simeq 0} e^{i q x} h(q) \\
    (-1)^x h_{\mathrm{b}}(x) &=  \sum_{q \simeq 0} e^{i q x} h(q + \pi).
 \end{split}
\end{align}
Note that in the case of $2k_{\mathrm{F}}=\pi$, the backward scattering leads to a real field $h_{\mathrm{b}}(x)$. 
Since $h_{\mathrm{f}}(f)$ and $h_{\mathrm{b}}(x)$ involve components of $h(q)$ which are different in $q$, their cross averages are zero and they can be considered as two independent random fields with essentially delta function correlations upon average. 
The fermionic field $\psi(x)=\sum_{k}e^{ikx}\hat{a}_{k}$ 
can be represented in terms of right- and left-movers~\cite{Giamarchi2004Book} 
\begin{align}
 \psi(x) = e^{i k_{\mathrm{F}} x} \psi_{\mathrm{R}}(x) 
    + e^{-i k_{\mathrm{F}} x} \psi_{\mathrm{L}}(x),
\end{align}
and the disorder term is recast into 
\begin{align}
 \hat{\mathcal{H}}_{\mathrm{dis}}=
   &-\int dx h_{\mathrm{f}}(x)
   [\psi_{\mathrm{R}}^{\dagger}(x) \psi_{\mathrm{R}}(x)
   + \psi_{\mathrm{L}}^{\dagger}(x)\psi_{\mathrm{L}}(x)]
\nonumber\\
 &- \int dx h_{\mathrm{b}}(x)
   [\psi_{\mathrm{R}}^{\dagger}(x) \psi_{\mathrm{L}}(x)
   + \psi_{\mathrm{L}}^{\dagger}(x)\psi_{\mathrm{R}}(x)],
\end{align}
by noting $h_{\mathrm{b}}^{*}(x)=h_{\mathrm{b}}(x)$. 
The current operator becomes
\begin{align}
 \hat{j} = v_{\mathrm{F}} 
   [\psi_{\mathrm{R}}^{\dagger}(x) \psi_{\mathrm{R}}(x)
     - \psi_{\mathrm{L}}^{\dagger}(x)\psi_{\mathrm{L}}(x)]
\end{align}
where $v_{\mathrm{F}}$ is the Fermi velocity.
The forward scattering part of the disorder can be eliminated by the gauge transformation~\cite{GiamarchiSchulz1988PRB}: 
\begin{align}
\begin{split}
 &\psi_{\mathrm{R}} \to
   e^{\frac{i}{v_\mathrm{F}} \int_{0}^{x} dy h_f(y)}
     \psi_{\mathrm{R}}(x)\\
 &\psi_{\mathrm{L}} \to
   e^{-\frac{i}{v_\mathrm{F}} \int_{0}^{x} dy h_f(y)}
     \psi_{\mathrm{L}}(x)
\end{split}
\end{align}
This transformation does not affect the current operator,
but changes the backscattering term to 
\begin{align}
 -\int dx h_{\mathrm{b}}(x)
   \big[e^{\frac{2 i}{v_\mathrm{F}} \int_{0}^{x} dy h_{\mathrm{f}}(y)}
   \psi_{\mathrm{R}}^{\dagger}(x) \psi_{\mathrm{L}}(x)
     + \mathrm{H.c.}\big]
\end{align}
One thus see that the backscattering part of disorder is replaced by a \textit{complex} field 
$\xi(x)=h_{\mathrm{b}}(x)e^{2 \frac{i}{v_\mathrm{F}} \int_{0}^{x} dy h_{\mathrm{f}}(y)}$ 
with correlations 
\begin{align}
 \begin{split}
   \overline{\xi(x)\xi(y)} &= 0, \\
   \overline{\xi(x)\xi^{*}(y)} &= D_{\mathrm{b}} \delta(x-y),
 \end{split}
\end{align}
where $D_{\mathrm{b}}\propto W^{2}$. 
This reflects the breaking of the particle-hole symmetry 
which is caused by one realization of the random chemical potential. 
Note that a system with perfect particle hole symmetry, 
such as a random bond system, would lead to a different fixed point, namely the random singlet phase~\cite{Fisher1994PRB}. 

We employ the usual representation of the fermion operators in terms of collective fields related to density and current, respectively~\cite{Giamarchi2004Book}, and connect them in the spin language [Eq.~\eqref{eq:spinham}]
to the two angles needed to dictate the direction of the spin vector 
\begin{align}
\begin{split}
 S_{l}^{+} &\simeq (-1)^{l} e^{i \theta(x)} 
   + e^{i \theta(x)} \cos(2\phi(x)), \\ 
 S_{l}^{z} &\simeq \frac{-\nabla\phi(x)}\pi 
   + (-1)^{l} \cos(2\phi(x)),
\end{split}
\end{align}
where $x$ is the position of $l$-th spin,
and the prefactors are omitted. 
Then the Hamiltonian is written as~\cite{GiamarchiSchulz1988PRB,Giamarchi2004Book}
\begin{align} 
 \hat{\mathcal{H}} 
   =& \frac1{2\pi} \int dx 
     \Big[u K (\nabla\theta(x))^{2} + \frac{u}{K} (\nabla\phi(x))^{2}\Big]
\nonumber\\
   &- \int dx [\xi(x) e^{i 2 \phi(x)} + \mathrm{H.c}]
\label{eq:hambos}
\end{align}
Here $u$ is the velocity of excitations 
($u=v_{\mathrm{F}}$ for the noninteracting case) 
and $K$ is the dimensionless parameter controlling the decay of correlations. 
We have neglected the irrelevant operator $\cos(4\phi(x))$ 
which appears for the special case of nonmagnetized spin chains, 
or equivalently half-filled fermion chains. 
This operator is irrelevant for $K>1/2$.

In this representation the current operator is the time derivative of the field $\phi(x)$~\cite{Giamarchi2004Book} 
and thus the conductivity Eq.~\eqref{eq:OptConduct} is simply related 
to the Green's function of the field $\phi(x)$ by 
\begin{align}
 \sigma(\omega)
   = -\frac{i2\omega}{\pi} \int dx \int_{0}^{\infty} dt 
   \braket{[\phi(x,t),\phi(0,0)]} e^{i\omega t- \epsilon t}
\end{align}
where $\sigma$ is in units of $e/h$ 
and $\epsilon=0^{+}$ is a convergence factor. 

The disorder term in Eq.~\eqref{eq:hambos} can be eliminated 
for a Gaussian disorder by using the replica trick \cite{GiamarchiSchulz1988PRB,Giamarchi2004Book}. 
This leads to an action of the form
\begin{align} 
 S =& \frac1{2\pi K} \sum_{\alpha} \int dx d\tau
   \Big[\frac{1}{u}(\partial_{\tau}\phi_{\alpha}(x))^2
     + u (\partial_{x}\phi_{\alpha}(x))^{2}\Big]
\nonumber\\ 
 &- D_{\mathrm{b}} \sum_{\alpha,\nu}
   \int dx d\tau_1 d\tau_2
     \cos[2\phi_{\alpha}(x,\tau_{1}) - 2\phi_{\nu}(x,\tau_{2})]
\label{eq:acdis}
\end{align}
where $\alpha,\nu = 1,2,\ldots,n$ are the replica indices 
and the limit $n\to 0$ must be taken. 

The disorder term is relevant for $K < 3/2$ even for an infinitesimal strength of the disorder $D_{\mathrm{b}}$. 
For $K > 3/2$ ($\Delta < -0.5$), there is a separatrix, with 
a transition of the Berezinskii-Kosterlitz-Thouless universality class 
between the localized and a delocalized TLL phase~\cite{GiamarchiSchulz1988PRB}.

The localization length can be captured by an RG evaluation of this problem. Far from the transition point $K=3/2$ the localization length behaves 
as~\cite{GiamarchiSchulz1988PRB}
\begin{align}
 \xi \propto \Big(\frac{1}{D_{\mathrm{b}}}\Big)^{\frac{1}{3-2K}}
\label{eq:locdep}
\end{align}
In addition to the RG method which can only give access to the physical quantities, either in the regime where the disorder 
is irrelevant or for scales smaller than the localization length, 
physical observables can be computed in the localized 
phase using a replica Gaussian variational approach~\cite{GiamarchiLeDoussal1996PRB}. This approximate approach leads to a low frequency behavior 
of the real part of the conductivity as
\begin{align}
    \sigma(\omega) \propto \omega^{2} \xi^{3}
\label{eq:condloclength}
\end{align}
where $\xi$ is the localization length. 
It does not include the logarithmic correction, 
which is known to occur for noninteracting particles 
$\sigma(\omega)\propto \omega^{2}(\ln\omega)^{2}$~\cite{Berezinskii1974JTEP,Gogolin1982PhysRep}. 
This is clearly an artifact of the variational approach 
and a logarithmic correction is also a priori expected for interacting systems
as suggested by semiclassical calculations $K \to 0$~\cite{Fogler2002PRL} 
and instanton expansions~\cite{Nattermann2003PRL,RosenowNattermann2006PRB}.

Equation~\eqref{eq:condloclength} implies that 
the pinning frequency $\omega_{\mathrm{p}}$ 
and the corresponding value of the dynamical conductivity 
$\sigma(\omega_{\mathrm{p}})$ scale as 
\begin{align}
 \omega_{\mathrm{p}}\propto\xi^{-1},\quad
 \sigma(\omega_{\mathrm{p}})\propto\xi.
\label{eq:peakscale}
\end{align}
in agreement with the predictions from the RG calculations as well. 
We use this scaling in comparison with the numerical results.


\begin{thebibliography}{41}%
\makeatletter
\providecommand \@ifxundefined [1]{%
 \@ifx{#1\undefined}
}%
\providecommand \@ifnum [1]{%
 \ifnum #1\expandafter \@firstoftwo
 \else \expandafter \@secondoftwo
 \fi
}%
\providecommand \@ifx [1]{%
 \ifx #1\expandafter \@firstoftwo
 \else \expandafter \@secondoftwo
 \fi
}%
\providecommand \natexlab [1]{#1}%
\providecommand \enquote  [1]{``#1''}%
\providecommand \bibnamefont  [1]{#1}%
\providecommand \bibfnamefont [1]{#1}%
\providecommand \citenamefont [1]{#1}%
\providecommand \href@noop [0]{\@secondoftwo}%
\providecommand \href [0]{\begingroup \@sanitize@url \@href}%
\providecommand \@href[1]{\@@startlink{#1}\@@href}%
\providecommand \@@href[1]{\endgroup#1\@@endlink}%
\providecommand \@sanitize@url [0]{\catcode `\\12\catcode `\$12\catcode
  `\&12\catcode `\#12\catcode `\^12\catcode `\_12\catcode `\%12\relax}%
\providecommand \@@startlink[1]{}%
\providecommand \@@endlink[0]{}%
\providecommand \url  [0]{\begingroup\@sanitize@url \@url }%
\providecommand \@url [1]{\endgroup\@href {#1}{\urlprefix }}%
\providecommand \urlprefix  [0]{URL }%
\providecommand \Eprint [0]{\href }%
\providecommand \doibase [0]{http://dx.doi.org/}%
\providecommand \selectlanguage [0]{\@gobble}%
\providecommand \bibinfo  [0]{\@secondoftwo}%
\providecommand \bibfield  [0]{\@secondoftwo}%
\providecommand \translation [1]{[#1]}%
\providecommand \BibitemOpen [0]{}%
\providecommand \bibitemStop [0]{}%
\providecommand \bibitemNoStop [0]{.\EOS\space}%
\providecommand \EOS [0]{\spacefactor3000\relax}%
\providecommand \BibitemShut  [1]{\csname bibitem#1\endcsname}%
\let\auto@bib@innerbib\@empty
%</preamble>
\bibitem [{\citenamefont {Anderson}(1958)}]{Anderson1958PR}%
  \BibitemOpen
  \bibfield  {author} {\bibinfo {author} {\bibfnamefont {P.~W.}\ \bibnamefont
  {Anderson}},\ }\href {\doibase 10.1103/PhysRev.109.1492} {\bibfield
  {journal} {\bibinfo  {journal} {Phys. Rev.}\ }\textbf {\bibinfo {volume}
  {109}},\ \bibinfo {pages} {1492} (\bibinfo {year} {1958})}\BibitemShut
  {NoStop}%
\bibitem [{\citenamefont {Sanchez-Palencia}\ \emph {et~al.}(2007)\citenamefont
  {Sanchez-Palencia}, \citenamefont {Cl\'ement}, \citenamefont {Lugan},
  \citenamefont {Bouyer}, \citenamefont {Shlyapnikov},\ and\ \citenamefont
  {Aspect}}]{Aspect_loc2007PRL}%
  \BibitemOpen
  \bibfield  {author} {\bibinfo {author} {\bibfnamefont {L.}~\bibnamefont
  {Sanchez-Palencia}}, \bibinfo {author} {\bibfnamefont {D.}~\bibnamefont
  {Cl\'ement}}, \bibinfo {author} {\bibfnamefont {P.}~\bibnamefont {Lugan}},
  \bibinfo {author} {\bibfnamefont {P.}~\bibnamefont {Bouyer}}, \bibinfo
  {author} {\bibfnamefont {G.~V.}\ \bibnamefont {Shlyapnikov}}, \ and\ \bibinfo
  {author} {\bibfnamefont {A.}~\bibnamefont {Aspect}},\ }\href {\doibase
  10.1103/PhysRevLett.98.210401} {\bibfield  {journal} {\bibinfo  {journal}
  {Phys. Rev. Lett.}\ }\textbf {\bibinfo {volume} {98}},\ \bibinfo {pages}
  {210401} (\bibinfo {year} {2007})}\BibitemShut {NoStop}%
\bibitem [{\citenamefont {Billy}\ \emph {et~al.}(2008)\citenamefont {Billy},
  \citenamefont {Josse}, \citenamefont {Zuo}, \citenamefont {Bernard},
  \citenamefont {Hambrecht}, \citenamefont {Lugan}, \citenamefont {Clement},
  \citenamefont {Sanchez-Palencia}, \citenamefont {Bouyer},\ and\ \citenamefont
  {Aspect}}]{Aspect_loc2008Nature}%
  \BibitemOpen
  \bibfield  {author} {\bibinfo {author} {\bibfnamefont {J.}~\bibnamefont
  {Billy}}, \bibinfo {author} {\bibfnamefont {V.}~\bibnamefont {Josse}},
  \bibinfo {author} {\bibfnamefont {Z.}~\bibnamefont {Zuo}}, \bibinfo {author}
  {\bibfnamefont {A.}~\bibnamefont {Bernard}}, \bibinfo {author} {\bibfnamefont
  {B.}~\bibnamefont {Hambrecht}}, \bibinfo {author} {\bibfnamefont
  {P.}~\bibnamefont {Lugan}}, \bibinfo {author} {\bibfnamefont
  {D.}~\bibnamefont {Clement}}, \bibinfo {author} {\bibfnamefont
  {L.}~\bibnamefont {Sanchez-Palencia}}, \bibinfo {author} {\bibfnamefont
  {P.}~\bibnamefont {Bouyer}}, \ and\ \bibinfo {author} {\bibfnamefont
  {A.}~\bibnamefont {Aspect}},\ }\href {\doibase 10.1038/nature07000}
  {\bibfield  {journal} {\bibinfo  {journal} {Nature (London)}\ }\textbf
  {\bibinfo {volume} {453}},\ \bibinfo {pages} {891} (\bibinfo {year}
  {2008})}\BibitemShut {NoStop}%
\bibitem [{\citenamefont {Fallani}\ \emph {et~al.}(2007)\citenamefont
  {Fallani}, \citenamefont {Lye}, \citenamefont {Guarrera}, \citenamefont
  {Fort},\ and\ \citenamefont {Inguscio}}]{Inguscio_loc2007PRL}%
  \BibitemOpen
  \bibfield  {author} {\bibinfo {author} {\bibfnamefont {L.}~\bibnamefont
  {Fallani}}, \bibinfo {author} {\bibfnamefont {J.~E.}\ \bibnamefont {Lye}},
  \bibinfo {author} {\bibfnamefont {V.}~\bibnamefont {Guarrera}}, \bibinfo
  {author} {\bibfnamefont {C.}~\bibnamefont {Fort}}, \ and\ \bibinfo {author}
  {\bibfnamefont {M.}~\bibnamefont {Inguscio}},\ }\href {\doibase
  10.1103/PhysRevLett.98.130404} {\bibfield  {journal} {\bibinfo  {journal}
  {Phys. Rev. Lett.}\ }\textbf {\bibinfo {volume} {98}},\ \bibinfo {pages}
  {130404} (\bibinfo {year} {2007})}\BibitemShut {NoStop}%
\bibitem [{\citenamefont {Roati}\ \emph {et~al.}(2008)\citenamefont {Roati},
  \citenamefont {D'Errico}, \citenamefont {Fallani}, \citenamefont {Fattori},
  \citenamefont {Fort}, \citenamefont {Zaccanti}, \citenamefont {Modugno},
  \citenamefont {Modugno},\ and\ \citenamefont
  {Inguscio}}]{Inguscio_loc2008Nature}%
  \BibitemOpen
  \bibfield  {author} {\bibinfo {author} {\bibfnamefont {G.}~\bibnamefont
  {Roati}}, \bibinfo {author} {\bibfnamefont {C.}~\bibnamefont {D'Errico}},
  \bibinfo {author} {\bibfnamefont {L.}~\bibnamefont {Fallani}}, \bibinfo
  {author} {\bibfnamefont {M.}~\bibnamefont {Fattori}}, \bibinfo {author}
  {\bibfnamefont {C.}~\bibnamefont {Fort}}, \bibinfo {author} {\bibfnamefont
  {M.}~\bibnamefont {Zaccanti}}, \bibinfo {author} {\bibfnamefont
  {G.}~\bibnamefont {Modugno}}, \bibinfo {author} {\bibfnamefont
  {M.}~\bibnamefont {Modugno}}, \ and\ \bibinfo {author} {\bibfnamefont
  {M.}~\bibnamefont {Inguscio}},\ }\href {\doibase 10.1038/nature07071}
  {\bibfield  {journal} {\bibinfo  {journal} {Nature (London)}\ }\textbf
  {\bibinfo {volume} {453}},\ \bibinfo {pages} {895} (\bibinfo {year}
  {2008})}\BibitemShut {NoStop}%
\bibitem [{\citenamefont {Filoche}\ and\ \citenamefont
  {Mayboroda}(2012)}]{Filoche_landscape2012PNAS}%
  \BibitemOpen
  \bibfield  {author} {\bibinfo {author} {\bibfnamefont {M.}~\bibnamefont
  {Filoche}}\ and\ \bibinfo {author} {\bibfnamefont {S.}~\bibnamefont
  {Mayboroda}},\ }\href {\doibase 10.1073/pnas.1120432109} {\bibfield
  {journal} {\bibinfo  {journal} {Proc. Natl. Acad. Sci.}\ }\textbf {\bibinfo
  {volume} {109}},\ \bibinfo {pages} {14761} (\bibinfo {year}
  {2012})}\BibitemShut {NoStop}%
\bibitem [{\citenamefont {Altshuler}\ \emph {et~al.}(1980)\citenamefont
  {Altshuler}, \citenamefont {Aronov},\ and\ \citenamefont
  {Lee}}]{AltshulerAronovLee1980PRL}%
  \BibitemOpen
  \bibfield  {author} {\bibinfo {author} {\bibfnamefont {B.~L.}\ \bibnamefont
  {Altshuler}}, \bibinfo {author} {\bibfnamefont {A.~G.}\ \bibnamefont
  {Aronov}}, \ and\ \bibinfo {author} {\bibfnamefont {P.~A.}\ \bibnamefont
  {Lee}},\ }\href {\doibase 10.1103/PhysRevLett.44.1288} {\bibfield  {journal}
  {\bibinfo  {journal} {Phys. Rev. Lett.}\ }\textbf {\bibinfo {volume} {44}},\
  \bibinfo {pages} {1288} (\bibinfo {year} {1980})}\BibitemShut {NoStop}%
\bibitem [{\citenamefont {Giamarchi}\ and\ \citenamefont
  {Schulz}(1988)}]{GiamarchiSchulz1988PRB}%
  \BibitemOpen
  \bibfield  {author} {\bibinfo {author} {\bibfnamefont {T.}~\bibnamefont
  {Giamarchi}}\ and\ \bibinfo {author} {\bibfnamefont {H.~J.}\ \bibnamefont
  {Schulz}},\ }\href {\doibase 10.1103/PhysRevB.37.325} {\bibfield  {journal}
  {\bibinfo  {journal} {Phys. Rev. B}\ }\textbf {\bibinfo {volume} {37}},\
  \bibinfo {pages} {325} (\bibinfo {year} {1988})}\BibitemShut {NoStop}%
\bibitem [{\citenamefont {Castellani}\ \emph {et~al.}(1984)\citenamefont
  {Castellani}, \citenamefont {Di~Castro}, \citenamefont {Lee},\ and\
  \citenamefont {Ma}}]{Castellani1984PRB}%
  \BibitemOpen
  \bibfield  {author} {\bibinfo {author} {\bibfnamefont {C.}~\bibnamefont
  {Castellani}}, \bibinfo {author} {\bibfnamefont {C.}~\bibnamefont
  {Di~Castro}}, \bibinfo {author} {\bibfnamefont {P.~A.}\ \bibnamefont {Lee}},
  \ and\ \bibinfo {author} {\bibfnamefont {M.}~\bibnamefont {Ma}},\ }\href
  {\doibase 10.1103/PhysRevB.30.527} {\bibfield  {journal} {\bibinfo  {journal}
  {Phys. Rev. B}\ }\textbf {\bibinfo {volume} {30}},\ \bibinfo {pages} {527}
  (\bibinfo {year} {1984})}\BibitemShut {NoStop}%
\bibitem [{\citenamefont {Finkel'stein}(1984)}]{Finkelstein1984ZPB}%
  \BibitemOpen
  \bibfield  {author} {\bibinfo {author} {\bibfnamefont {A.~M.}\ \bibnamefont
  {Finkel'stein}},\ }\href {\doibase 10.1007/BF01304171} {\bibfield  {journal}
  {\bibinfo  {journal} {Z. Phys. B Condens. Matter}\ }\textbf {\bibinfo
  {volume} {56}},\ \bibinfo {pages} {189} (\bibinfo {year} {1984})}\BibitemShut
  {NoStop}%
\bibitem [{\citenamefont {Belitz}\ and\ \citenamefont
  {Kirkpatrick}(1994)}]{BelitzKirkpatrick1994RMP}%
  \BibitemOpen
  \bibfield  {author} {\bibinfo {author} {\bibfnamefont {D.}~\bibnamefont
  {Belitz}}\ and\ \bibinfo {author} {\bibfnamefont {T.~R.}\ \bibnamefont
  {Kirkpatrick}},\ }\href {\doibase 10.1103/RevModPhys.66.261} {\bibfield
  {journal} {\bibinfo  {journal} {Rev. Mod. Phys.}\ }\textbf {\bibinfo {volume}
  {66}},\ \bibinfo {pages} {261} (\bibinfo {year} {1994})}\BibitemShut
  {NoStop}%
\bibitem [{\citenamefont {Gornyi}\ \emph {et~al.}(2005)\citenamefont {Gornyi},
  \citenamefont {Mirlin},\ and\ \citenamefont
  {Polyakov}}]{GornyiMirlinPolyakov2005PRL}%
  \BibitemOpen
  \bibfield  {author} {\bibinfo {author} {\bibfnamefont {I.~V.}\ \bibnamefont
  {Gornyi}}, \bibinfo {author} {\bibfnamefont {A.~D.}\ \bibnamefont {Mirlin}},
  \ and\ \bibinfo {author} {\bibfnamefont {D.~G.}\ \bibnamefont {Polyakov}},\
  }\href {\doibase 10.1103/PhysRevLett.95.206603} {\bibfield  {journal}
  {\bibinfo  {journal} {Phys. Rev. Lett.}\ }\textbf {\bibinfo {volume} {95}},\
  \bibinfo {pages} {206603} (\bibinfo {year} {2005})}\BibitemShut {NoStop}%
\bibitem [{\citenamefont {Basko}\ \emph {et~al.}(2007)\citenamefont {Basko},
  \citenamefont {Aleiner},\ and\ \citenamefont
  {Altshuler}}]{BaskoAleinerAltshuler2007PRB}%
  \BibitemOpen
  \bibfield  {author} {\bibinfo {author} {\bibfnamefont {D.~M.}\ \bibnamefont
  {Basko}}, \bibinfo {author} {\bibfnamefont {I.~L.}\ \bibnamefont {Aleiner}},
  \ and\ \bibinfo {author} {\bibfnamefont {B.~L.}\ \bibnamefont {Altshuler}},\
  }\href {\doibase 10.1103/PhysRevB.76.052203} {\bibfield  {journal} {\bibinfo
  {journal} {Phys. Rev. B}\ }\textbf {\bibinfo {volume} {76}},\ \bibinfo
  {pages} {052203} (\bibinfo {year} {2007})}\BibitemShut {NoStop}%
\bibitem [{\citenamefont {Oganesyan}\ and\ \citenamefont
  {Huse}(2007)}]{OganesyanHuse2007PRB}%
  \BibitemOpen
  \bibfield  {author} {\bibinfo {author} {\bibfnamefont {V.}~\bibnamefont
  {Oganesyan}}\ and\ \bibinfo {author} {\bibfnamefont {D.~A.}\ \bibnamefont
  {Huse}},\ }\href {\doibase 10.1103/PhysRevB.75.155111} {\bibfield  {journal}
  {\bibinfo  {journal} {Phys. Rev. B}\ }\textbf {\bibinfo {volume} {75}},\
  \bibinfo {pages} {155111} (\bibinfo {year} {2007})}\BibitemShut {NoStop}%
\bibitem [{\citenamefont {Abanin}\ \emph {et~al.}(2019)\citenamefont {Abanin},
  \citenamefont {Altman}, \citenamefont {Bloch},\ and\ \citenamefont
  {Serbyn}}]{Abanin_mbl2019RMP}%
  \BibitemOpen
  \bibfield  {author} {\bibinfo {author} {\bibfnamefont {D.~A.}\ \bibnamefont
  {Abanin}}, \bibinfo {author} {\bibfnamefont {E.}~\bibnamefont {Altman}},
  \bibinfo {author} {\bibfnamefont {I.}~\bibnamefont {Bloch}}, \ and\ \bibinfo
  {author} {\bibfnamefont {M.}~\bibnamefont {Serbyn}},\ }\href {\doibase
  10.1103/RevModPhys.91.021001} {\bibfield  {journal} {\bibinfo  {journal}
  {Rev. Mod. Phys.}\ }\textbf {\bibinfo {volume} {91}},\ \bibinfo {pages}
  {021001} (\bibinfo {year} {2019})}\BibitemShut {NoStop}%
\bibitem [{\citenamefont {Giamarchi}(2004)}]{Giamarchi2004Book}%
  \BibitemOpen
  \bibfield  {author} {\bibinfo {author} {\bibfnamefont {T.}~\bibnamefont
  {Giamarchi}},\ }\href@noop {} {\emph {\bibinfo {title} {Quantum physics in
  one dimension}}}\ (\bibinfo  {publisher} {Oxford university press, Oxford},\
  \bibinfo {year} {2004})\BibitemShut {NoStop}%
\bibitem [{\citenamefont {Fisher}\ \emph {et~al.}(1989)\citenamefont {Fisher},
  \citenamefont {Weichman}, \citenamefont {Grinstein},\ and\ \citenamefont
  {Fisher}}]{Fisher_bosonloc1989PRB}%
  \BibitemOpen
  \bibfield  {author} {\bibinfo {author} {\bibfnamefont {M.~P.~A.}\
  \bibnamefont {Fisher}}, \bibinfo {author} {\bibfnamefont {P.~B.}\
  \bibnamefont {Weichman}}, \bibinfo {author} {\bibfnamefont {G.}~\bibnamefont
  {Grinstein}}, \ and\ \bibinfo {author} {\bibfnamefont {D.~S.}\ \bibnamefont
  {Fisher}},\ }\href {\doibase 10.1103/PhysRevB.40.546} {\bibfield  {journal}
  {\bibinfo  {journal} {Phys. Rev. B}\ }\textbf {\bibinfo {volume} {40}},\
  \bibinfo {pages} {546} (\bibinfo {year} {1989})}\BibitemShut {NoStop}%
\bibitem [{\citenamefont {Tanzi}\ \emph {et~al.}(2013)\citenamefont {Tanzi},
  \citenamefont {Lucioni}, \citenamefont {Chaudhuri}, \citenamefont {Gori},
  \citenamefont {Kumar}, \citenamefont {D'Errico}, \citenamefont {Inguscio},\
  and\ \citenamefont {Modugno}}]{Modugno_bg2013PRL}%
  \BibitemOpen
  \bibfield  {author} {\bibinfo {author} {\bibfnamefont {L.}~\bibnamefont
  {Tanzi}}, \bibinfo {author} {\bibfnamefont {E.}~\bibnamefont {Lucioni}},
  \bibinfo {author} {\bibfnamefont {S.}~\bibnamefont {Chaudhuri}}, \bibinfo
  {author} {\bibfnamefont {L.}~\bibnamefont {Gori}}, \bibinfo {author}
  {\bibfnamefont {A.}~\bibnamefont {Kumar}}, \bibinfo {author} {\bibfnamefont
  {C.}~\bibnamefont {D'Errico}}, \bibinfo {author} {\bibfnamefont
  {M.}~\bibnamefont {Inguscio}}, \ and\ \bibinfo {author} {\bibfnamefont
  {G.}~\bibnamefont {Modugno}},\ }\href {\doibase
  10.1103/PhysRevLett.111.115301} {\bibfield  {journal} {\bibinfo  {journal}
  {Phys. Rev. Lett.}\ }\textbf {\bibinfo {volume} {111}},\ \bibinfo {pages}
  {115301} (\bibinfo {year} {2013})}\BibitemShut {NoStop}%
\bibitem [{\citenamefont {D'Errico}\ \emph {et~al.}(2014)\citenamefont
  {D'Errico}, \citenamefont {Lucioni}, \citenamefont {Tanzi}, \citenamefont
  {Gori}, \citenamefont {Roux}, \citenamefont {McCulloch}, \citenamefont
  {Giamarchi}, \citenamefont {Inguscio},\ and\ \citenamefont
  {Modugno}}]{Modugno_bg2014PRL}%
  \BibitemOpen
  \bibfield  {author} {\bibinfo {author} {\bibfnamefont {C.}~\bibnamefont
  {D'Errico}}, \bibinfo {author} {\bibfnamefont {E.}~\bibnamefont {Lucioni}},
  \bibinfo {author} {\bibfnamefont {L.}~\bibnamefont {Tanzi}}, \bibinfo
  {author} {\bibfnamefont {L.}~\bibnamefont {Gori}}, \bibinfo {author}
  {\bibfnamefont {G.}~\bibnamefont {Roux}}, \bibinfo {author} {\bibfnamefont
  {I.~P.}\ \bibnamefont {McCulloch}}, \bibinfo {author} {\bibfnamefont
  {T.}~\bibnamefont {Giamarchi}}, \bibinfo {author} {\bibfnamefont
  {M.}~\bibnamefont {Inguscio}}, \ and\ \bibinfo {author} {\bibfnamefont
  {G.}~\bibnamefont {Modugno}},\ }\href {\doibase
  10.1103/PhysRevLett.113.095301} {\bibfield  {journal} {\bibinfo  {journal}
  {Phys. Rev. Lett.}\ }\textbf {\bibinfo {volume} {113}},\ \bibinfo {pages}
  {095301} (\bibinfo {year} {2014})}\BibitemShut {NoStop}%
\bibitem [{\citenamefont {Mott}(1970)}]{Mott1970PhilMag}%
  \BibitemOpen
  \bibfield  {author} {\bibinfo {author} {\bibfnamefont {N.~F.}\ \bibnamefont
  {Mott}},\ }\href {\doibase 10.1080/14786437008228147} {\bibfield  {journal}
  {\bibinfo  {journal} {Phil. Mag.}\ }\textbf {\bibinfo {volume} {22}},\
  \bibinfo {pages} {7} (\bibinfo {year} {1970})}\BibitemShut {NoStop}%
\bibitem [{\citenamefont {Nattermann}\ \emph {et~al.}(2003)\citenamefont
  {Nattermann}, \citenamefont {Giamarchi},\ and\ \citenamefont
  {Le~Doussal}}]{Nattermann2003PRL}%
  \BibitemOpen
  \bibfield  {author} {\bibinfo {author} {\bibfnamefont {T.}~\bibnamefont
  {Nattermann}}, \bibinfo {author} {\bibfnamefont {T.}~\bibnamefont
  {Giamarchi}}, \ and\ \bibinfo {author} {\bibfnamefont {P.}~\bibnamefont
  {Le~Doussal}},\ }\href {\doibase 10.1103/PhysRevLett.91.056603} {\bibfield
  {journal} {\bibinfo  {journal} {Phys. Rev. Lett.}\ }\textbf {\bibinfo
  {volume} {91}},\ \bibinfo {pages} {056603} (\bibinfo {year}
  {2003})}\BibitemShut {NoStop}%
\bibitem [{\citenamefont {Aleiner}\ \emph {et~al.}(2010)\citenamefont
  {Aleiner}, \citenamefont {Altshuler},\ and\ \citenamefont
  {Shlyapnikov}}]{AleinerAltshulerShlyapnikov2010NatPhys}%
  \BibitemOpen
  \bibfield  {author} {\bibinfo {author} {\bibfnamefont {I.~L.}\ \bibnamefont
  {Aleiner}}, \bibinfo {author} {\bibfnamefont {B.~L.}\ \bibnamefont
  {Altshuler}}, \ and\ \bibinfo {author} {\bibfnamefont {G.~V.}\ \bibnamefont
  {Shlyapnikov}},\ }\href {\doibase 10.1038/nphys1758} {\bibfield  {journal}
  {\bibinfo  {journal} {Nat. Phys.}\ }\textbf {\bibinfo {volume} {6}},\
  \bibinfo {pages} {900} (\bibinfo {year} {2010})}\BibitemShut {NoStop}%
\bibitem [{\citenamefont {Berezinskii}(1974)}]{Berezinskii1974JTEP}%
  \BibitemOpen
  \bibfield  {author} {\bibinfo {author} {\bibfnamefont {V.~L.}\ \bibnamefont
  {Berezinskii}},\ }\href
  {http://www.jetp.ac.ru/cgi-bin/index/e/38/3/p620?a=list} {\bibfield
  {journal} {\bibinfo  {journal} {JETP}\ }\textbf {\bibinfo {volume} {38}},\
  \bibinfo {pages} {620} (\bibinfo {year} {1974})}\BibitemShut {NoStop}%
\bibitem [{\citenamefont {Gogolin}\ and\ \citenamefont
  {Melnikov}(1978)}]{Gogolin1978PSS}%
  \BibitemOpen
  \bibfield  {author} {\bibinfo {author} {\bibfnamefont {A.~A.}\ \bibnamefont
  {Gogolin}}\ and\ \bibinfo {author} {\bibfnamefont {V.~I.}\ \bibnamefont
  {Melnikov}},\ }\href {\doibase 10.1002/pssb.2220880202} {\bibfield  {journal}
  {\bibinfo  {journal} {phys. stat. sol. (b)}\ }\textbf {\bibinfo {volume}
  {88}},\ \bibinfo {pages} {377} (\bibinfo {year} {1978})}\BibitemShut
  {NoStop}%
\bibitem [{\citenamefont {Gogolin}(1982)}]{Gogolin1982PhysRep}%
  \BibitemOpen
  \bibfield  {author} {\bibinfo {author} {\bibfnamefont {A.~A.}\ \bibnamefont
  {Gogolin}},\ }\href {\doibase https://doi.org/10.1016/0370-1573(82)90069-2}
  {\bibfield  {journal} {\bibinfo  {journal} {Phys. Rep.}\ }\textbf {\bibinfo
  {volume} {86}},\ \bibinfo {pages} {1} (\bibinfo {year} {1982})}\BibitemShut
  {NoStop}%
\bibitem [{\citenamefont {Giamarchi}\ and\ \citenamefont
  {Le~Doussal}(1996)}]{GiamarchiLeDoussal1996PRB}%
  \BibitemOpen
  \bibfield  {author} {\bibinfo {author} {\bibfnamefont {T.}~\bibnamefont
  {Giamarchi}}\ and\ \bibinfo {author} {\bibfnamefont {P.}~\bibnamefont
  {Le~Doussal}},\ }\href {\doibase 10.1103/PhysRevB.53.15206} {\bibfield
  {journal} {\bibinfo  {journal} {Phys. Rev. B}\ }\textbf {\bibinfo {volume}
  {53}},\ \bibinfo {pages} {15206} (\bibinfo {year} {1996})}\BibitemShut
  {NoStop}%
\bibitem [{\citenamefont {Tokuno}\ and\ \citenamefont
  {Giamarchi}(2011)}]{TokunoGiamarchi2011PRL}%
  \BibitemOpen
  \bibfield  {author} {\bibinfo {author} {\bibfnamefont {A.}~\bibnamefont
  {Tokuno}}\ and\ \bibinfo {author} {\bibfnamefont {T.}~\bibnamefont
  {Giamarchi}},\ }\href {\doibase 10.1103/PhysRevLett.106.205301} {\bibfield
  {journal} {\bibinfo  {journal} {Phys. Rev. Lett.}\ }\textbf {\bibinfo
  {volume} {106}},\ \bibinfo {pages} {205301} (\bibinfo {year}
  {2011})}\BibitemShut {NoStop}%
\bibitem [{\citenamefont {Wu}\ \emph {et~al.}(2015)\citenamefont {Wu},
  \citenamefont {Taylor},\ and\ \citenamefont
  {Zaremba}}]{Wu_shakeconduct2015EPL}%
  \BibitemOpen
  \bibfield  {author} {\bibinfo {author} {\bibfnamefont {Z.}~\bibnamefont
  {Wu}}, \bibinfo {author} {\bibfnamefont {E.}~\bibnamefont {Taylor}}, \ and\
  \bibinfo {author} {\bibfnamefont {E.}~\bibnamefont {Zaremba}},\ }\href
  {\doibase 10.1209/0295-5075/110/26002} {\bibfield  {journal} {\bibinfo
  {journal} {Europhys. Lett.}\ }\textbf {\bibinfo {volume} {110}},\ \bibinfo
  {pages} {26002} (\bibinfo {year} {2015})}\BibitemShut {NoStop}%
\bibitem [{\citenamefont {Anderson}\ \emph {et~al.}(2019)\citenamefont
  {Anderson}, \citenamefont {Wang}, \citenamefont {Xu}, \citenamefont {Venu},
  \citenamefont {Trotzky}, \citenamefont {Chevy},\ and\ \citenamefont
  {Thywissen}}]{Anderson_optlattconduct2019PRL}%
  \BibitemOpen
  \bibfield  {author} {\bibinfo {author} {\bibfnamefont {R.}~\bibnamefont
  {Anderson}}, \bibinfo {author} {\bibfnamefont {F.}~\bibnamefont {Wang}},
  \bibinfo {author} {\bibfnamefont {P.}~\bibnamefont {Xu}}, \bibinfo {author}
  {\bibfnamefont {V.}~\bibnamefont {Venu}}, \bibinfo {author} {\bibfnamefont
  {S.}~\bibnamefont {Trotzky}}, \bibinfo {author} {\bibfnamefont
  {F.}~\bibnamefont {Chevy}}, \ and\ \bibinfo {author} {\bibfnamefont {J.~H.}\
  \bibnamefont {Thywissen}},\ }\href {\doibase 10.1103/PhysRevLett.122.153602}
  {\bibfield  {journal} {\bibinfo  {journal} {Phys. Rev. Lett.}\ }\textbf
  {\bibinfo {volume} {122}},\ \bibinfo {pages} {153602} (\bibinfo {year}
  {2019})}\BibitemShut {NoStop}%
\bibitem [{\citenamefont {Orso}\ \emph {et~al.}(2009)\citenamefont {Orso},
  \citenamefont {Iucci}, \citenamefont {Cazalilla},\ and\ \citenamefont
  {Giamarchi}}]{Orso_optlattdisorder2009PRA}%
  \BibitemOpen
  \bibfield  {author} {\bibinfo {author} {\bibfnamefont {G.}~\bibnamefont
  {Orso}}, \bibinfo {author} {\bibfnamefont {A.}~\bibnamefont {Iucci}},
  \bibinfo {author} {\bibfnamefont {M.~A.}\ \bibnamefont {Cazalilla}}, \ and\
  \bibinfo {author} {\bibfnamefont {T.}~\bibnamefont {Giamarchi}},\ }\href
  {\doibase 10.1103/PhysRevA.80.033625} {\bibfield  {journal} {\bibinfo
  {journal} {Phys. Rev. A}\ }\textbf {\bibinfo {volume} {80}},\ \bibinfo
  {pages} {033625} (\bibinfo {year} {2009})}\BibitemShut {NoStop}%
\bibitem [{\citenamefont {White}(1992)}]{White1992PRL}%
  \BibitemOpen
  \bibfield  {author} {\bibinfo {author} {\bibfnamefont {S.~R.}\ \bibnamefont
  {White}},\ }\href {\doibase 10.1103/PhysRevLett.69.2863} {\bibfield
  {journal} {\bibinfo  {journal} {Phys. Rev. Lett.}\ }\textbf {\bibinfo
  {volume} {69}},\ \bibinfo {pages} {2863} (\bibinfo {year}
  {1992})}\BibitemShut {NoStop}%
\bibitem [{\citenamefont {Schollw\"ock}(2011)}]{Schollwock2011AnnPhys}%
  \BibitemOpen
  \bibfield  {author} {\bibinfo {author} {\bibfnamefont {U.}~\bibnamefont
  {Schollw\"ock}},\ }\href {\doibase 10.1016/j.aop.2010.09.012} {\bibfield
  {journal} {\bibinfo  {journal} {Ann. Phys.}\ }\textbf {\bibinfo {volume}
  {326}},\ \bibinfo {pages} {96} (\bibinfo {year} {2011})}\BibitemShut
  {NoStop}%
\bibitem [{\citenamefont {Cazalilla}\ and\ \citenamefont
  {Marston}(2002)}]{CazalillaMarston2002PRL}%
  \BibitemOpen
  \bibfield  {author} {\bibinfo {author} {\bibfnamefont {M.~A.}\ \bibnamefont
  {Cazalilla}}\ and\ \bibinfo {author} {\bibfnamefont {J.~B.}\ \bibnamefont
  {Marston}},\ }\href {\doibase 10.1103/PhysRevLett.88.256403} {\bibfield
  {journal} {\bibinfo  {journal} {Phys. Rev. Lett.}\ }\textbf {\bibinfo
  {volume} {88}},\ \bibinfo {pages} {256403} (\bibinfo {year}
  {2002})}\BibitemShut {NoStop}%
\bibitem [{\citenamefont {Vidal}(2004)}]{Vidal_tebd2004PRL}%
  \BibitemOpen
  \bibfield  {author} {\bibinfo {author} {\bibfnamefont {G.}~\bibnamefont
  {Vidal}},\ }\href {\doibase 10.1103/PhysRevLett.93.040502} {\bibfield
  {journal} {\bibinfo  {journal} {Phys. Rev. Lett.}\ }\textbf {\bibinfo
  {volume} {93}},\ \bibinfo {pages} {040502} (\bibinfo {year}
  {2004})}\BibitemShut {NoStop}%
\bibitem [{\citenamefont {Hallberg}(2006)}]{Hallberg2006AdvPhys}%
  \BibitemOpen
  \bibfield  {author} {\bibinfo {author} {\bibfnamefont {K.~A.}\ \bibnamefont
  {Hallberg}},\ }\href {\doibase 10.1080/00018730600766432} {\bibfield
  {journal} {\bibinfo  {journal} {Adv. Phys.}\ }\textbf {\bibinfo {volume}
  {55}},\ \bibinfo {pages} {477} (\bibinfo {year} {2006})}\BibitemShut
  {NoStop}%
\bibitem [{\citenamefont {Holzner}\ \emph {et~al.}(2011)\citenamefont
  {Holzner}, \citenamefont {Weichselbaum}, \citenamefont {McCulloch},
  \citenamefont {Schollw\"ock},\ and\ \citenamefont {von
  Delft}}]{Holzner2011PRB}%
  \BibitemOpen
  \bibfield  {author} {\bibinfo {author} {\bibfnamefont {A.}~\bibnamefont
  {Holzner}}, \bibinfo {author} {\bibfnamefont {A.}~\bibnamefont
  {Weichselbaum}}, \bibinfo {author} {\bibfnamefont {I.~P.}\ \bibnamefont
  {McCulloch}}, \bibinfo {author} {\bibfnamefont {U.}~\bibnamefont
  {Schollw\"ock}}, \ and\ \bibinfo {author} {\bibfnamefont {J.}~\bibnamefont
  {von Delft}},\ }\href {\doibase 10.1103/PhysRevB.83.195115} {\bibfield
  {journal} {\bibinfo  {journal} {Phys. Rev. B}\ }\textbf {\bibinfo {volume}
  {83}},\ \bibinfo {pages} {195115} (\bibinfo {year} {2011})}\BibitemShut
  {NoStop}%
\bibitem [{\citenamefont {Wei\ss{}e}\ \emph {et~al.}(2006)\citenamefont
  {Wei\ss{}e}, \citenamefont {Wellein}, \citenamefont {Alvermann},\ and\
  \citenamefont {Fehske}}]{Weisse2006RMP}%
  \BibitemOpen
  \bibfield  {author} {\bibinfo {author} {\bibfnamefont {A.}~\bibnamefont
  {Wei\ss{}e}}, \bibinfo {author} {\bibfnamefont {G.}~\bibnamefont {Wellein}},
  \bibinfo {author} {\bibfnamefont {A.}~\bibnamefont {Alvermann}}, \ and\
  \bibinfo {author} {\bibfnamefont {H.}~\bibnamefont {Fehske}},\ }\href
  {\doibase 10.1103/RevModPhys.78.275} {\bibfield  {journal} {\bibinfo
  {journal} {Rev. Mod. Phys.}\ }\textbf {\bibinfo {volume} {78}},\ \bibinfo
  {pages} {275} (\bibinfo {year} {2006})}\BibitemShut {NoStop}%
\bibitem [{\citenamefont {Abrikosov}\ and\ \citenamefont
  {Ryzhkin}(1978)}]{AbrikosovRyzhkin1978AdvPhys}%
  \BibitemOpen
  \bibfield  {author} {\bibinfo {author} {\bibfnamefont {A.~A.}\ \bibnamefont
  {Abrikosov}}\ and\ \bibinfo {author} {\bibfnamefont {I.~A.}\ \bibnamefont
  {Ryzhkin}},\ }\href {\doibase 10.1080/00018737800101364} {\bibfield
  {journal} {\bibinfo  {journal} {Adv. Phys.}\ }\textbf {\bibinfo {volume}
  {27}},\ \bibinfo {pages} {147} (\bibinfo {year} {1978})}\BibitemShut
  {NoStop}%
\bibitem [{\citenamefont {Fogler}(2002)}]{Fogler2002PRL}%
  \BibitemOpen
  \bibfield  {author} {\bibinfo {author} {\bibfnamefont {M.~M.}\ \bibnamefont
  {Fogler}},\ }\href {\doibase 10.1103/PhysRevLett.88.186402} {\bibfield
  {journal} {\bibinfo  {journal} {Phys. Rev. Lett.}\ }\textbf {\bibinfo
  {volume} {88}},\ \bibinfo {pages} {186402} (\bibinfo {year}
  {2002})}\BibitemShut {NoStop}%
\bibitem [{\citenamefont {Rosenow}\ and\ \citenamefont
  {Nattermann}(2006)}]{RosenowNattermann2006PRB}%
  \BibitemOpen
  \bibfield  {author} {\bibinfo {author} {\bibfnamefont {B.}~\bibnamefont
  {Rosenow}}\ and\ \bibinfo {author} {\bibfnamefont {T.}~\bibnamefont
  {Nattermann}},\ }\href {\doibase 10.1103/PhysRevB.73.085103} {\bibfield
  {journal} {\bibinfo  {journal} {Phys. Rev. B}\ }\textbf {\bibinfo {volume}
  {73}},\ \bibinfo {pages} {085103} (\bibinfo {year} {2006})}\BibitemShut
  {NoStop}%
\bibitem [{\citenamefont {Fisher}(1994)}]{Fisher1994PRB}%
  \BibitemOpen
  \bibfield  {author} {\bibinfo {author} {\bibfnamefont {D.~S.}\ \bibnamefont
  {Fisher}},\ }\href {\doibase 10.1103/PhysRevB.50.3799} {\bibfield  {journal}
  {\bibinfo  {journal} {Phys. Rev. B}\ }\textbf {\bibinfo {volume} {50}},\
  \bibinfo {pages} {3799} (\bibinfo {year} {1994})}\BibitemShut {NoStop}%
\end{thebibliography}
\end{document}